\documentclass[article,12pt]{amsart}
\usepackage{amsfonts}
\usepackage{amsmath}
\usepackage{graphicx}
\usepackage{color}
\usepackage{chngcntr}
\usepackage{float}

\numberwithin{equation}{section}

\theoremstyle{plain}

\theoremstyle{remark}

\newcommand{\cN}{\mathcal{N}}

\newcommand{\dE}{\mathbb{E}}
\newcommand{\dN}{\mathbb{N}}
\newcommand{\dP}{\mathbb{P}}
\newcommand{\dR}{\mathbb{R}}
\newcommand{\dV}{\mathbb{V}}

\newcommand{\dd}{\mathrm{d}}
\newcommand{\de}{\mathrm{e}}

\newcommand{\hsp}{\hspace{0.5cm}}
\newcommand{\wt}{\widetilde}

\newcommand{\veps}{\varepsilon}

\def\build#1_#2^#3{\mathrel{\mathop{\kern 0pt#1}\limits_{#2}^{#3}}}

\def\videbox{\mathbin{\vbox{\hrule\hbox{\vrule height1ex \kern.5em
\vrule height1ex}\hrule}}}

\counterwithin{figure}{section}
\counterwithin{table}{section}

\email{Frederic.Proia@univ-angers.fr}
\email{Alix.Pernet@angers.inra.fr}
\email{Tatiana.Thouroude@angers.inra.fr}
\email{Gilles.Michel@angers.inra.fr}
\email{Jeremy.Clotault@univ-angers.fr}


\begin{document}

\title[On the characterization of flowering curves]
{On the characterization of flowering curves using Gaussian mixture models
\vspace{2ex}}
\author[F. Pro\"ia]{Fr\'ed\'eric Pro\"ia}
\address{Laboratoire Angevin de REcherche en MAth\'ematiques -- UMR 6093, Universit\'e d'Angers, D\'epartement de math\'ematiques, Facult\'e des Sciences, 2 Boulevard Lavoisier, 49045 Angers cedex, France. \vspace{-0.6cm}}

\author[A. Pernet]{Alix Pernet}
\author[T. Thouroude]{Tatiana Thouroude}
\author[G. Michel]{Gilles Michel}
\address{Institut de Recherche en Horticulture et Semences -- UMR 1345, INRA, SFR 4207 QuaSaV, 42 rue Georges Morel, 49071 Beaucouz\'e cedex, France.}

\author[J. Clotault]{J\'er\'emy Clotault}
\address{Institut de Recherche en Horticulture et Semences -- UMR 1345, Universit\'e d'Angers, SFR 4207 QuaSaV, 42 rue Georges Morel, 49071 Beaucouz\'e cedex, France.}
\thanks{}

\begin{abstract}
In this paper, we develop a statistical methodology applied to the characterization of flowering curves using Gaussian mixture models. Our study relies on a set of rosebushes flowering data, and Gaussian mixture models are mainly used to quantify the reblooming properties of each one. In this regard, we also suggest our own selection criterion to take into account the lack of symmetry of most of the flowering curves. Three classes are created on the basis of a principal component analysis conducted on a set of reblooming indicators, and a subclassification is made using a longitudinal $k$--means algorithm which also highlights the role played by the precocity of the flowering. In this way, we obtain an overview of the correlations between the features we decided to retain on each curve. In particular, results suggest the lack of correlation between reblooming and flowering precocity. The pertinent indicators obtained in this study will be a first step towards the comprehension of the environmental and genetic control of these biological processes.
\end{abstract}

\maketitle


\section{Introduction and Motivations}
\label{SecIntro}

As it is explained by Putterill \textit{et al.} \cite{Putterill04}, matching the flowering period with the best climatic conditions is a crucial step for wild plants to obtain a high fertility rate. In agriculture, the amount of seeds and fruits produced by plants is directly related to their ability to produce a great number of flowers, hence flowering is extremely important for high-yields crops. For ornamental plants, obtaining a large number of flowers over the longest period of the year is an important breeding objective.

\smallskip

Plants present a large diversity of flowering patterns between taxa and suitable parameters  are necessary to summarize these flowering profiles. Flowering curves, counting the number of flowers observed for a plant at regular time intervals, can be obtained from field scorings. Statistical methods have been rarely used to efficiently describe and compare flowering curves. As an example, regression curves have been used to fit flowering curves, but only for \textit{once-flowering} plants whose curve shape skewed from normality (see Clark and Thompson \cite{ClarkThompson11}). Especially in horticulture, annual flowering curves are sometimes much more complex. \textit{Reblooming} -- or \textit{recurrent flowering} -- plants are able to flower and fructify several times over the year. Such plants are found among several ornamental species, like irises, hydrangeas, daylilies or roses, but also in fruit-producing species like strawberry or raspberry plants.

\smallskip

For roses (\textit{Rosa} sp. or genus \textit{Rosa}), flowering traits are particularly important, either for cut or garden roses. In Occident, the nineteenth century represents a golden age for rosebush breeding. It involved the creation of many cultivars, with the introduction of new traits in created hybrids, as explained by Oghina-Pavie \cite{OghinaPavie15}. Very early in this century, breeding activities have aimed at obtaining earlier -- or later -- flowering cultivars to increase the range of flowering periods (Oghina-Pavie, pers. comm.). Later in this century, the reblooming trait became the most important trait in rosebush breeding. Current modern roses result from crosses between reblooming Chinese roses and once-flowering European roses, obtained during the nineteenth century, according to Wylie \cite{Wylie54}. By the number of created cultivars and by the diversification of flowering profiles, rosebush genetic resources of the nineteenth century are probably the base of the most interesting methodologies to be developed in characterizing flowering curves.

\begin{figure}[H]
\centering
\includegraphics[width=15cm]{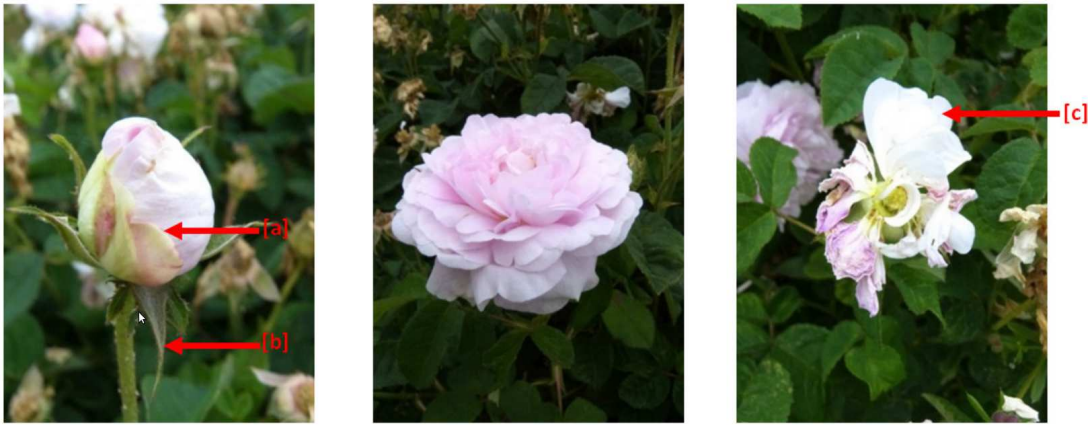}~\medskip
\caption{\scriptsize{Flowering stages for a rose (photo credit: Ballerie, 2012). On the left, [a] shows the detachment of one petal from the bud and [b] shows the detachment of one sepal from the bud. On the middle, a standard state of open flower with no withered petal is depicted. On the right, [c] points out the last non-withered petal. Were considered all flowers whose developmental stage lied between the left and right pictures.}}
\label{FigExFlow}
\end{figure}

The biological sample analyzed in this article is composed of 329 exploitable flowering curves obtained in 2012 in the rose garden ``Loubert'' (Les Rosiers-sur-Loire, France). The studied genotypes were predominantly bred during the nineteenth century. For each genotype, the number of open flowers was counted almost each week between May, 10th and November, 15th. For the most widespread case, were considered as open flowers developmental stages between these two following stages: (1) flower bud with at least one sepal detached from petals and at least one petal detached from the others (except for simple flowers, having five petals) and (2) flower with at least one petal which remains with original aspect and colour (see Figure \ref{FigExFlow} above). The plant shape (sphere, cylinder or cone), circumference and height were measured for calculation of flower density (number of flowers per m$^2$). The mean number of flowers within an inflorescence was also counted. The dataset originally contained some irregularities and missing values, and different temporal lags between consecutive measures. All these issues have been carefully dealt with by the authors, but occasional presence of residual artificial values cannot be excluded.

\smallskip

In rosebush, the contemporary works of Iwata \textit{et al.} \cite{Iwata12} and Kawamura \textit{et al.} \cite{Kawamura15} have highlighted the fact that flowering process is tightly linked to the branching process of the plant. In once-flowering cultivars, inflorescences are produced in the spring by the development of shoots from axillary buds of shoots from the previous year. Later in the year, new indeterminate shoots are produced and remain vegetative (having no flower). Inflorescence will develop the year after from axillary buds, borne by these vegetative shoots. In reblooming cultivars, either axillary buds will give inflorescence, or new determinate shoots terminated by an inflorescence will emerge successively from older shoots. Therefore, the best way to characterize rose flowering profile would be to differentiate the number of flowers produced by each shoot developed along the year. As an illustration of decomposition of the flowering shoot by shoot, Durand \textit{et al.} \cite{Durand13} tried to model biennial bearing in apple trees. For a large sample of elderly rosebushes with many shoots like the one studied in this article, this represents a huge and laborious work. Statistical methods are therefore needed to characterize flowering profiles (flowering date, flowering intensity, reblooming magnitude) from flowering curves obtained by counting flowers along the year in the whole plant. As for the characterization of reblooming, it is especially challenging to distinguish a long unique flowering period from several partially overlapping ones, corresponding to successive floral initiations.

\smallskip

Mixture models have for a long time been popular in life sciences, especially in biology and genetics, in fact since the seminal works of Pearson in the late nineteenth century. We guide the reader to the far from exhaustive mixtures applications to biology by Hale and Knott \cite{HaleyKnott92} in 1992, Lynch and Walsh \cite{LynchWalsh98} in 1998, Detilleux and Leroy \cite{DetilleuxLeroy00} in 2000, Boettcher \textit{et al.} \cite{Boettcher05} in 2005, Choi \textit{et al.} \cite{ChoiQinGhosh10} in 2010, Shekofteh \textit{et al.} \cite{Shekofteh15} in 2015, and references therein. The ability of a Gaussian mixture to split an apparently chaotic whole phenomenon into simple components and to highlight hidden structures was our main motivation to apply such models to flowering curves. Indeed, a rosebush is made of stems among which one may start to bloom while another starts to loose its flowers. On a whole set of branches, the resulting phenomenon is not suitably explained from a deterministic approach consisting in counting all variations from one week to the other. On the contrary, the \textit{waves} mechanism of Gaussian mixture models seems to form a relevant alternative, as we will see in the Appendix. The paper is organized as follows. Sections \ref{SecAppli} and \ref{SecClassif} are devoted to the statistical tools that we intend to customize and to their applications on our dataset, respectively for the characterization and the classification of the flowering curves. In particular, a theoretical background is supplied, when necessary. Some concluding remarks are given in Section \ref{SecConclu} and a schematic example is provided in the Appendix, to justify our choice of Gaussian mixture models.

\newpage


\section{An Application of GMM to Flowering Curves}
\label{SecAppli}

This section is devoted to the application of the \textit{Gaussian Mixture Model} -- shortened from now on \textit{GMM} -- on a set of flowering curves according to a statistical methodology that we will detail. Firstly, we need to supply a short theoretical background about GMM (see \cite{McLachlanBasford88}--\cite{McLachlanPeel00} for more details).
 
\subsection{The Gaussian Mixture Model}

Consider a set $(X_1, \hdots, X_{n})$ of $n$ real-valued random variables that we want to divide in $k$ classes. For all $1 \leq i \leq n$, we denote by $Z_{i}$ the latent random variable in $\{1, \hdots, k\}$ corresponding to the class of $X_{i}$. We suppose that $(Z_1, \hdots, Z_{n})$ are independent and have the same distribution as a random variable $Z$ such that, for all $1 \leq j \leq k$,
$$
\dP(Z = j) = \pi_{j}.
$$
In addition, we suppose that for all $1 \leq i \leq n$, the random variable $X_{i}\, \vert\, \{ Z_{i} = j \}$ has a $\cN(\mu_{j}, \sigma^2_{j})$ distribution and accordingly, $\pi_{j}$ stands for the proportion of the class $j$ in the whole population. For all $x \in \dR$, the distribution of the mixture is
$$
f_{\textnormal{GM}}(x) = \sum_{j=1}^{k} \pi_{j}\, f(x\, \vert\, \mu_{j}, \sigma^2_{j})
$$
where $f(\cdot\, \vert\, \mu_{j}, \sigma^2_{j})$ is the Gaussian distribution function with parameters $\mu_{j}$ and $\sigma^2_{j}$. If we consider that, given a subdivision in $k$ classes (that is, conditionally on the latent variables), the sequence $(X_1, \hdots, X_{n})$ is made of independent variables, then the (incomplete) log-likelihood for a set $\beta_{k} = (\pi_1, \hdots, \pi_{k}, \mu_1, \hdots, \mu_{k}, \sigma^2_1, \hdots, \sigma^2_{k})$ of $3k$ parameters is given for any observation $x = (x_1, \hdots, x_{n}) \in \dR^{n}$ by
\begin{equation}
\label{LogLik}
\ln \ell_{\textnormal{GM}}(x\, \vert\, \beta_{k}) = \sum_{i=1}^{n} \ln\left( \sum_{j=1}^{k} \pi_{j}\, f(x_{i}\, \vert\, \mu_{j}, \sigma^2_{j}) \right) = \sum_{i=1}^{n} \ln f_{\textnormal{GM}}(x_{i}).
\end{equation}
The classic approach (see e.g. \cite{Day69}--\cite{XuJordan96}) to estimate the $3k-1$ parameters, considering the relation $\pi_1 + \hdots + \pi_{k} = 1$, is to run the so-called \textit{Expectation-Maximisation algorithm} \cite{DempsterLairdRubin77} to maximize the above log-likelihood. The resulting estimator $\wt{\beta}_{k}$ is finally used to classify the observations \text{via} the Bayes' theorem. Namely,
$$
\wt{\dP}(Z_{i} = j\, \vert\, X_{i} = x_{i}) = \frac{\wt{\pi}_{j}\, f(x_{i}\, \vert\, \wt{\mu}_{j}, \wt{\sigma}^{\, 2}_{j})}{\sum_{\ell=1}^{k} \wt{\pi}_{\ell}\, f(x_{i}\, \vert\, \wt{\mu}_{\ell}, \wt{\sigma}^{\, 2}_{\ell})}
$$
usually leading to the posterior classification rule given, for all $1 \leq i \leq n$, by
$$
\wt{Z}_{i} ~ = ~ \arg \max_{1\, \leq\, j\, \leq\, k} \wt{\dP}(Z_{i} = j\, \vert\, X_{i} = x_{i}).
$$
For a given number of classes $k$, the \textit{Bayesian information criterion} \cite{Schwarz78} is
\begin{equation}
\label{BIC}
\textnormal{BIC}(k) = -2 \ln \ell_{\textnormal{GM}}(x\, \vert\, \wt{\beta}_{k}) + (3k - 1) \ln n
\end{equation}
where the log-likelihood is given in \eqref{LogLik}. It is then a natural solution to select
$$
\wt{k} ~ = ~ \arg \min_{1\, \leq\, k\, \leq\, k_{m}} \textnormal{BIC}(k)
$$
for an arbitrary upper bound $k_{m}$. 

\begin{figure}[H]
\centering
\includegraphics[width=14cm]{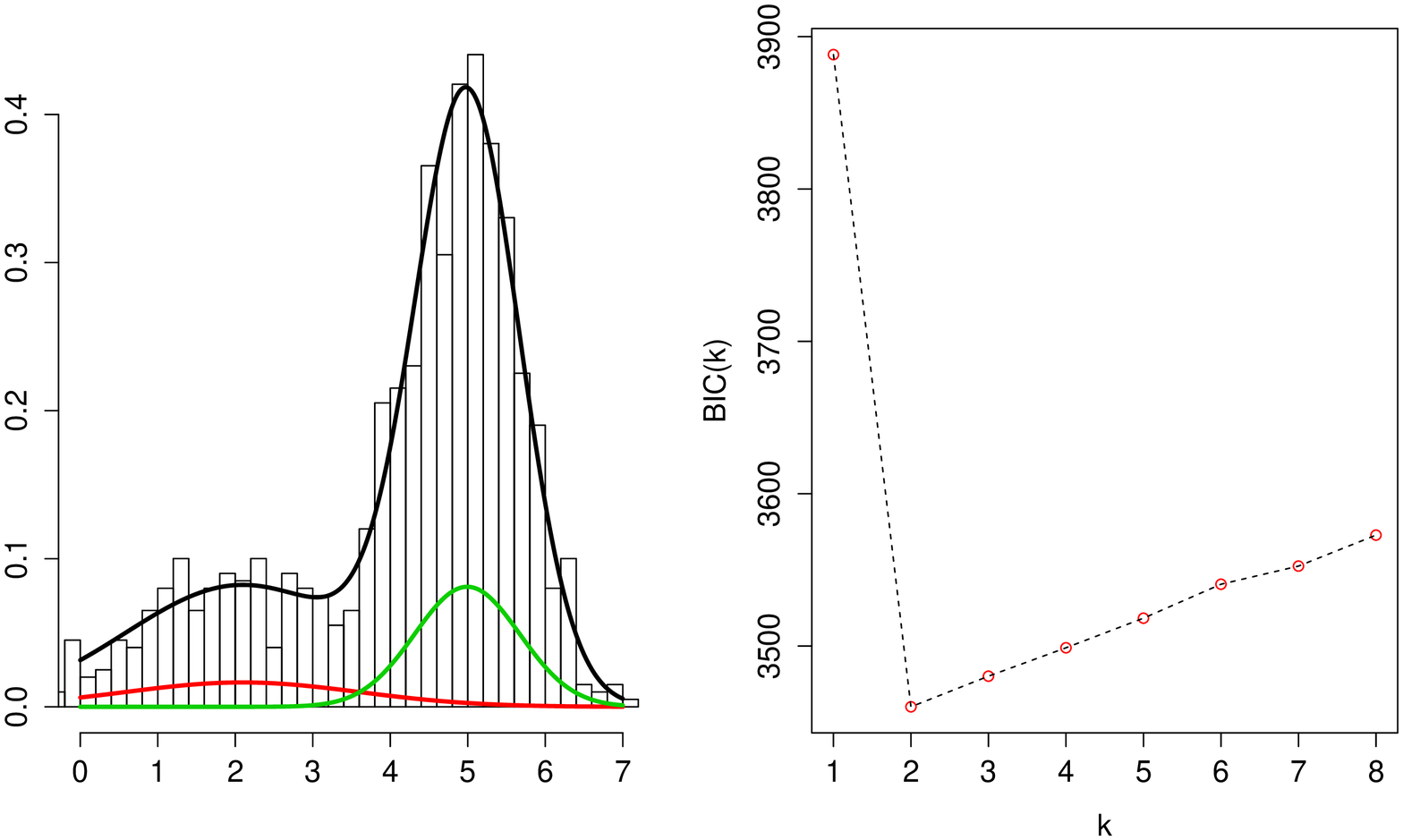}
\caption{\scriptsize{Example of GMM on simulated data according to $k=2$, $\pi=(0.3,0.7)$, $\mu=(2,5)$, $\sigma^2=(2,0.5)$. The coloured curves are the Gaussian components whereas the black one is the resulting mixture. The evolution of BIC is shown alongside for $k_{m}=8$.}}
\label{FigGMM1}
\end{figure}

\begin{figure}[H]
\centering
\includegraphics[width=14cm]{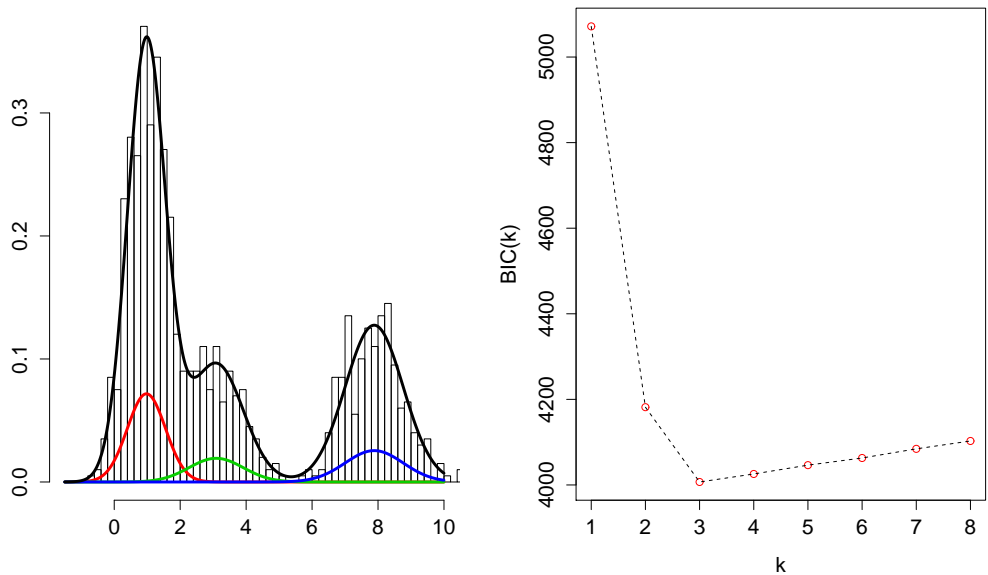}
\caption{\scriptsize{Example of GMM on simulated data according to $k=3$, $\pi=(0.5,0.2,0.3)$, $\mu=(1,3,8)$, $\sigma^2=(0.3,1,0.8)$. The coloured curves are the Gaussian components whereas the black one is the resulting mixture. The evolution of BIC is shown alongside for $k_{m}=8$.}}
\label{FigGMM2}
\end{figure}

\noindent In the following, the \texttt{R} library \texttt{mclust} \cite{RmclustMan}--\cite{FraleyRaftery02} is used to run the GMM procedures. On Figures \ref{FigGMM1}--\ref{FigGMM2} above, two examples of GMM are represented on simulated data with $n=1000$. The coloured curves stand for the Gaussian components whereas the black one is the resulting mixture. The evolution of BIC is shown alongside for $k_{m}=8$, leading to $\wt{k}=2$ and $\wt{k}=3$, respectively. A learning set is available on which we assume that the phenotypic behavior in terms of blooming is well-known. Working on these learning curves will help us to evaluate some parameters, such as $\alpha$ in the penalization that we will introduce. To apply a GMM on a flowering curve, we suggest to handle the curve as a probability distribution function and to simulate a sample in accordance with. In this context, the fitted mixture model characterizes the temporal probability distribution of the amount of flowers on the rosebush, along the year.
 
\subsection{The flowering curve as a distribution function}

First, we need to precise that what we call \textit{a flowering curve} is a discontinuous set of measures. The continuous line between all points that is represented on the figures throughout the study is only a visual tool, it is never technically used. In particular, the lack of information between each measure was precisely our main motivation to use the step interpolation function that we are going to describe. Let us consider the observed path $(y_{i,1}, \hdots, y_{i,d}) \in \dR^{d}$ associated with the $i$--th curve of the dataset, and $(t_{i,1}, \hdots, t_{i,d}) \in \dN^{d}$ corresponding to the instants of measure. For all $1 \leq \ell \leq d$, we build the step function $g_{i}$ according to
\begin{equation}
\label{StepFunc}
g_{i}(x) = y_{i,\ell} \hsp \text{for } x \in \left] t_{i,\ell} - \frac{t_{i,\ell} - t_{i,\ell-1}}{2}\, ;\, t_{i,\ell} + \frac{t_{i,\ell+1} - t_{i,\ell}}{2} \right]
\end{equation}
with the convention that $t_{i,0} = t_{i,1}-1$ and that $t_{i,d+1} = t_{i,d}+1$. On Figures \ref{FigDens1} and \ref{FigDens2} below, some examples are provided from the dataset. The blue lines are the flowering curves and the magenta rectangulations stand for the associated step functions $g_{i}$. Hence, the step function $\wt{g}_{i}$ defined for all $t_{i,0} \leq x \leq t_{i,d+1}$ by 
$$
\wt{g}_{i}(x) = \frac{g_{i}(x)}{A_{i}} \hsp \text{where} \hsp A_{i} = \int_{t_{i,0}}^{t_{i,d+1}} g_{i}(x)\, \dd x = \frac{1}{2}\, \sum_{\ell=1}^{d} y_{i,\ell}\, (t_{i,\ell+1} - t_{i,\ell-1}),
$$
can be seen as a probability distribution function. A GMM applied on a sample $(\wt{Y}_{i,1}, \hdots, \wt{Y}_{i,n_{s}})$ of $n_{s}$ independent random variables distributed according to $\wt{g}_{i}$ provides a tool to characterize the recurrent blooming of the $i$--th rosebush of the dataset. The smoothing effect on the chaotic behavior of most of the flowering curves is a substantial improvement compared to the deterministic strategy consisting in counting all changes of variations: the induced \textit{waves} mechanism of GMM seems somehow more adapted to the environmental and genetic reality of the plant, as it is shown in the schematic example provided in the Appendix. Through this illustration, we intend to highlight that the biological process has to be explained by the superimposition of hidden waves of flowering, and not simply by the evaluation of changes in the number of flowers over time. A weather effect can also involve a locally chaotic behavior, for example rain and wind are likely to produce a sudden fall of petals leading to biased measurements. Indeed, the counting process does not distinguish the ability of the rosebush to produce flowers from their resistance to time and weather.

\begin{figure}[H]
\centering
\includegraphics[width=14cm]{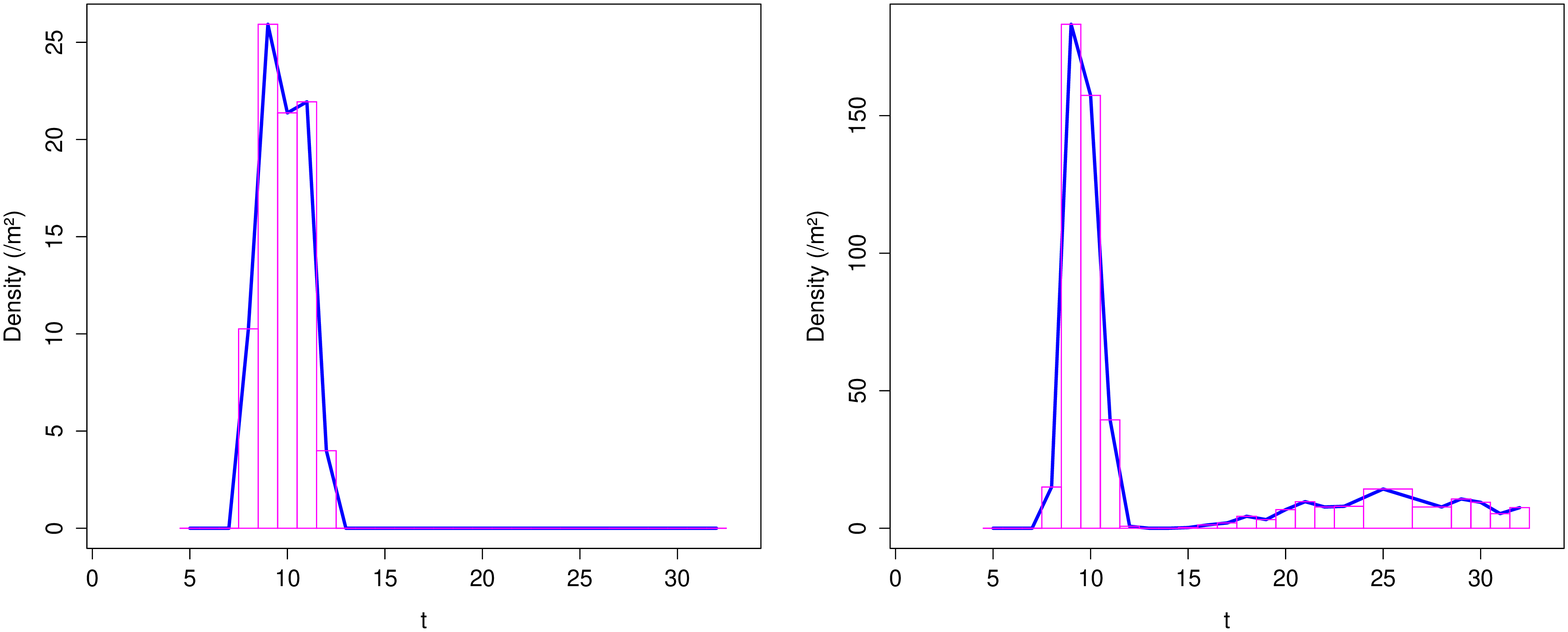}
\caption{\scriptsize{Examples of rectangulations (in magenta) of two flowering curves (in blue) using step functions.}}
\label{FigDens1}
\end{figure}

\begin{figure}[H]
\centering
\includegraphics[width=14cm]{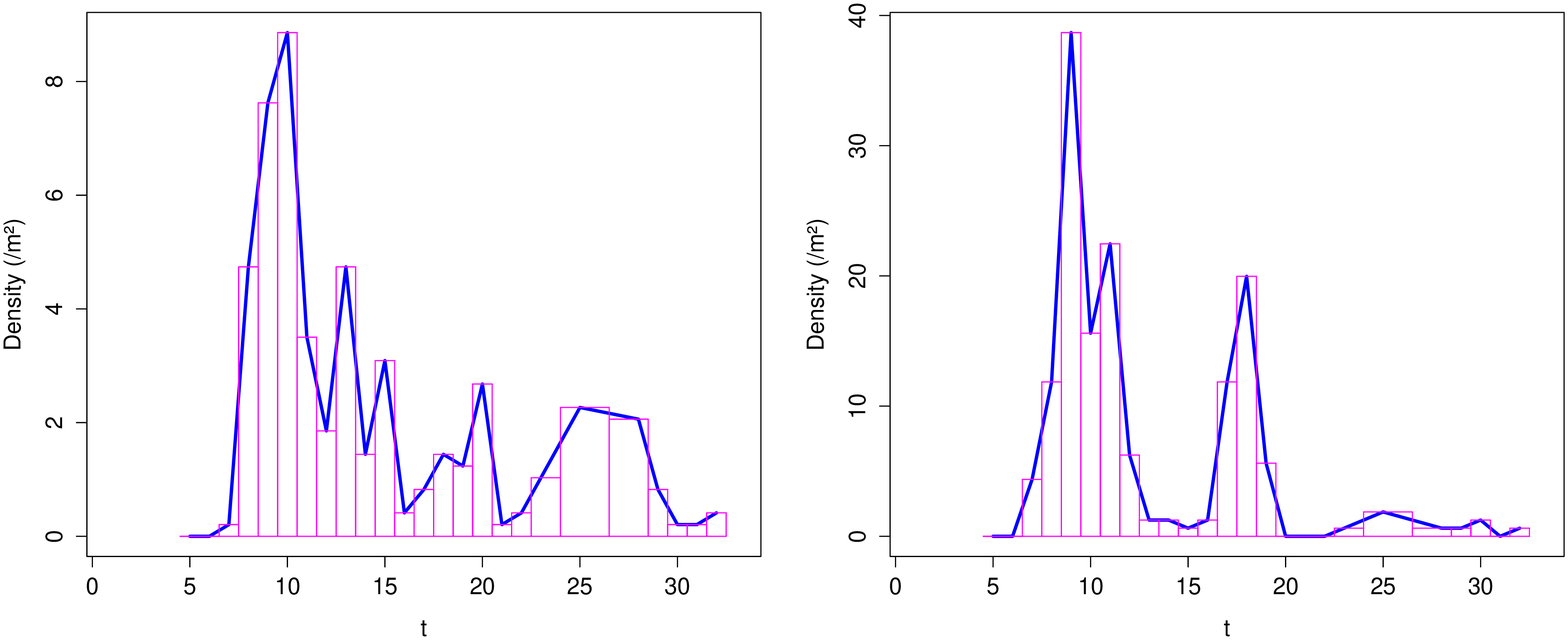}
\caption{\scriptsize{Examples of rectangulations (in magenta) of two flowering curves (in blue) using step functions.}}
\label{FigDens2}
\end{figure}

\subsection{An adapted criterion}

\begin{figure}[H]
\centering
\includegraphics[width=12cm]{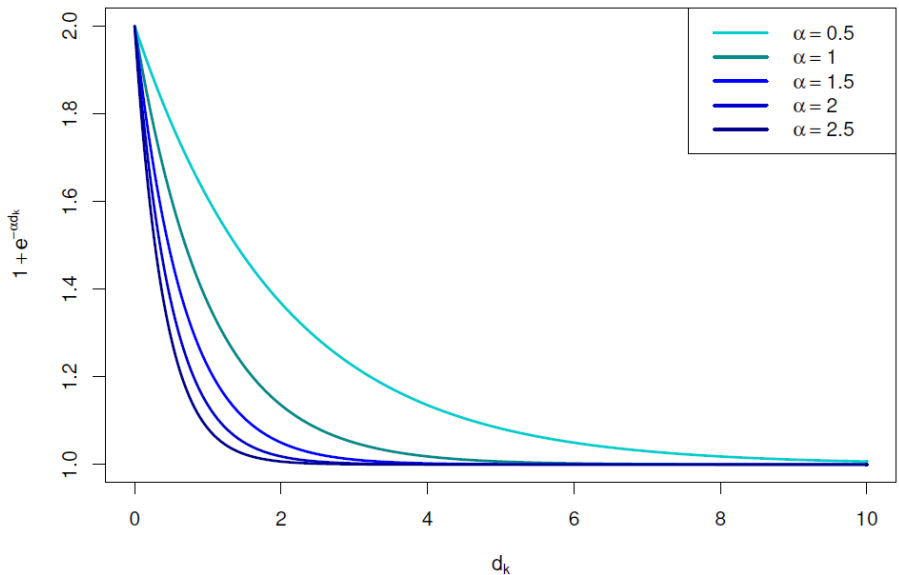}~\bigskip
\caption{\scriptsize{Evolution of the penalization coefficient $(1 + \de^{-\alpha\, d_{k}})$ according to $d_{k}$, for different values of $\alpha$.}}
\label{FigPen}
\end{figure}

The main issue arising in the GMM procedure is that, due to environment and biological effects, most often the generated sample fails in the usual statistical testing procedures for Gaussianity (such as Shapiro's) on the learning curves having a unique peak of blooming (clearly perceptible on the curves represented throughout the paper). This phenomenon is also observed in \cite{ClarkThompson11} and references within, though to a far lesser extent. The resulting asymmetry may lead to inappropriate conclusions from the GMM algorithm, in particular when some clusters become very close to each over to take into account all irregularities of the curve. To reduce this phenomenon, we suggest to use a selection criterion penalizing the $3k-1$ parameters to estimate (via the ordinary BIC), but also the smallest difference between the estimated means of the $k$ Gaussian components. Using the same notations as in the definition of BIC in \eqref{BIC} given an estimation on $k$ clusters, let us define
\begin{equation}
\label{BICm}
\textnormal{BIC}^{\,*}(k) = \big( c + \textnormal{BIC}(k) \big) \left( 1 + \de^{-\alpha\, d_{k}} \right)
\end{equation}
where $\alpha, c \in \dR^{+}$, $d_1 = +\infty$ and for $k \geq 2$,
\begin{equation}
\label{DifMin}
d_{k} = \min_{\underset{(j_1 \neq j_2)}{1\, \leq\, j_1,\, j_2\, \leq\, k}} \vert\, \wt{\mu}_{j_1} - \wt{\mu}_{j_2} \vert.
\end{equation}
Figure \ref{FigPen} shows the evolution of the penalization coefficient according to $d_{k}$, for $\alpha \in \{0.5, 1, 1.5, 2, 2.5 \}$. As we can see, this coefficient aims to sharply penalize any model where the smallest difference between the estimated means becomes less than 1.5, on the whole. According to us, the probability is higher that such a situation corresponds to a lack of symmetry of the current flowering.

\begin{figure}[H]
\centering
\includegraphics[width=16cm]{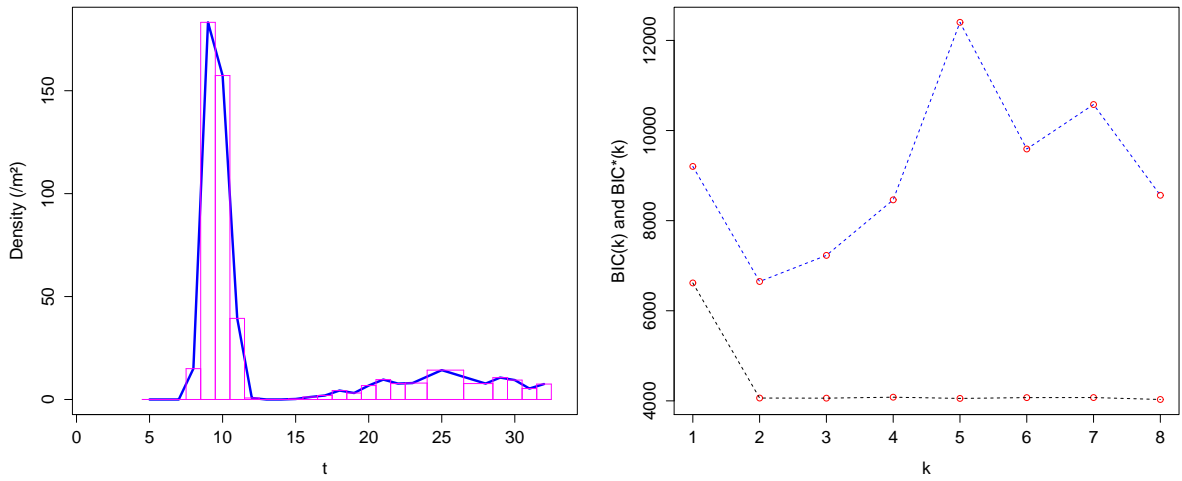}
\caption{\scriptsize{Example of flowering curve with its rectangulation and evolution of the associated BIC (in black) and $\textnormal{BIC}^{*}$ (in blue) from GMM estimation.}}
\label{FigExBICCurves1}
\end{figure}

\begin{figure}[H]
\centering
\includegraphics[width=16cm]{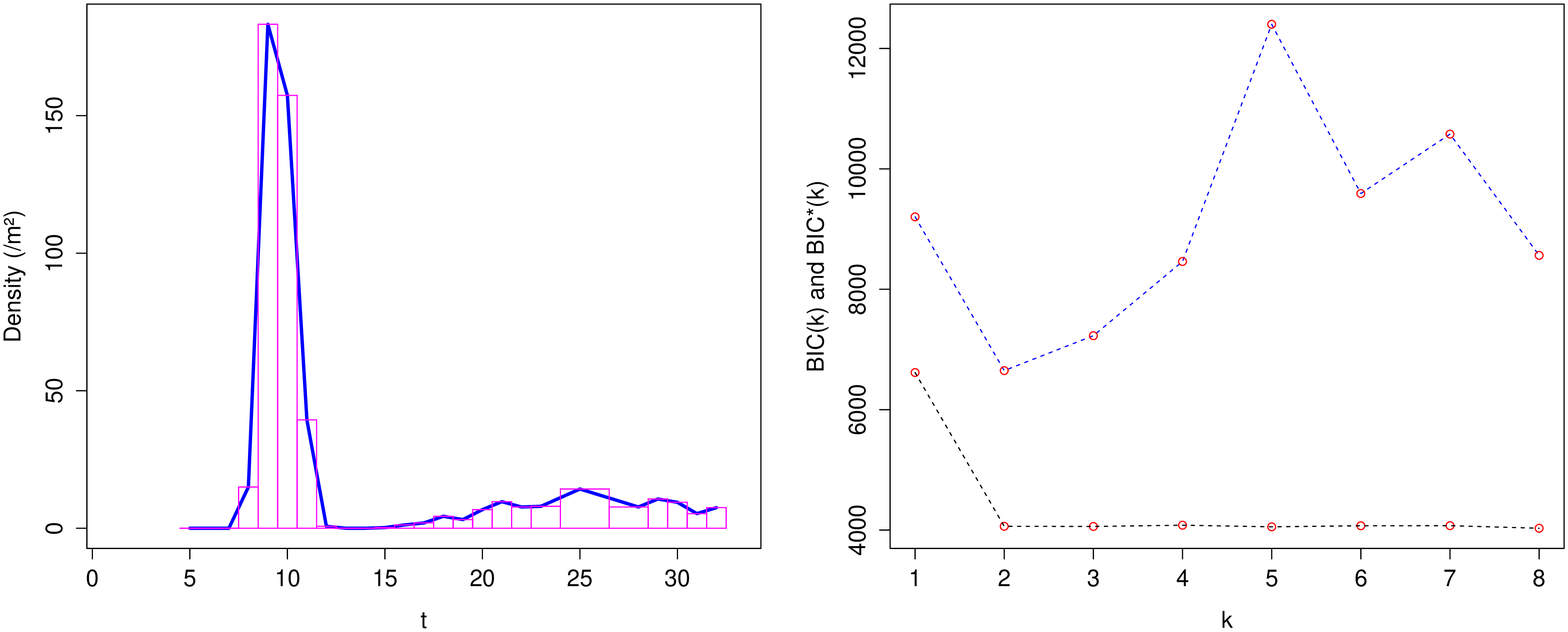}
\caption{\scriptsize{Example of flowering curve with its rectangulation and evolution of the associated BIC (in black) and $\textnormal{BIC}^{*}$ (in blue) from GMM estimation.}}
\label{FigExBICCurves2}
\end{figure}

\begin{figure}[H]
\centering
\includegraphics[width=16cm]{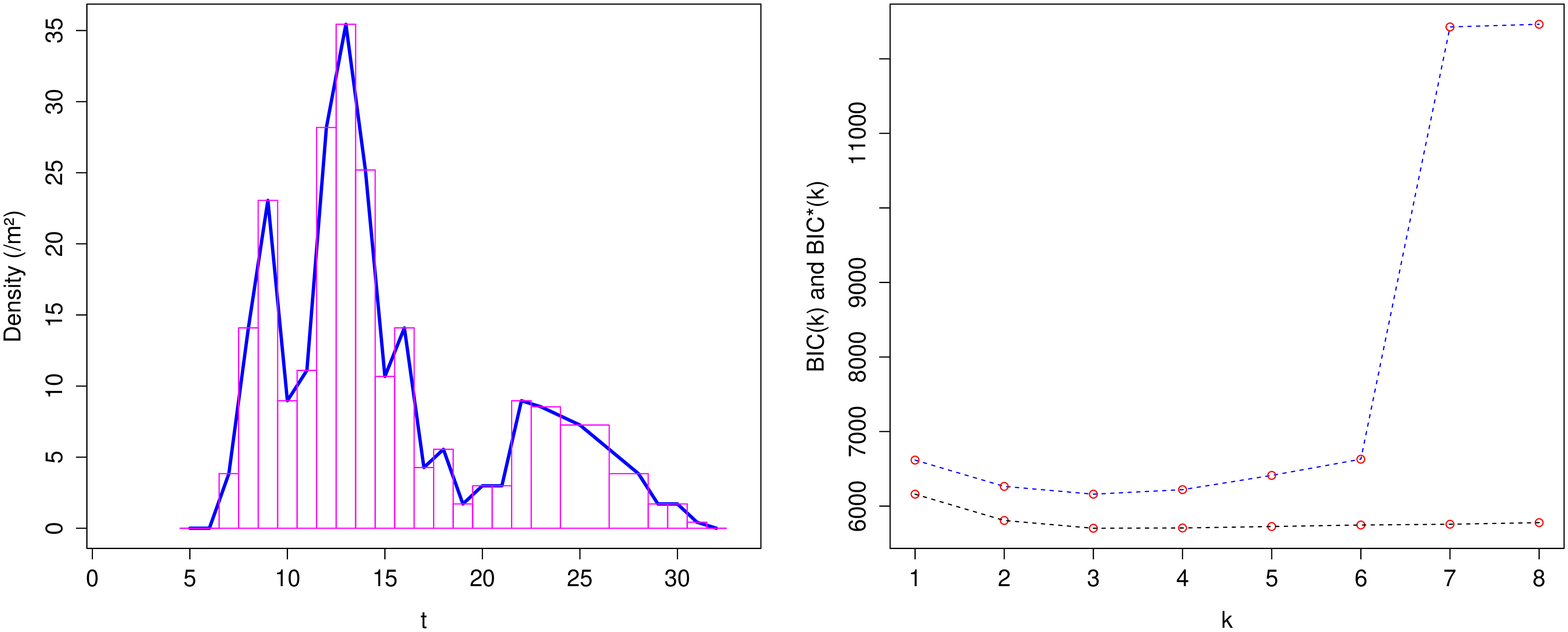}
\caption{\scriptsize{Example of flowering curve with its rectangulation and evolution of the associated BIC (in black) and $\textnormal{BIC}^{*}$ (in blue) from GMM estimation.}}
\label{FigExBICCurves3}
\end{figure}

\begin{figure}[H]
\centering
\includegraphics[width=16cm]{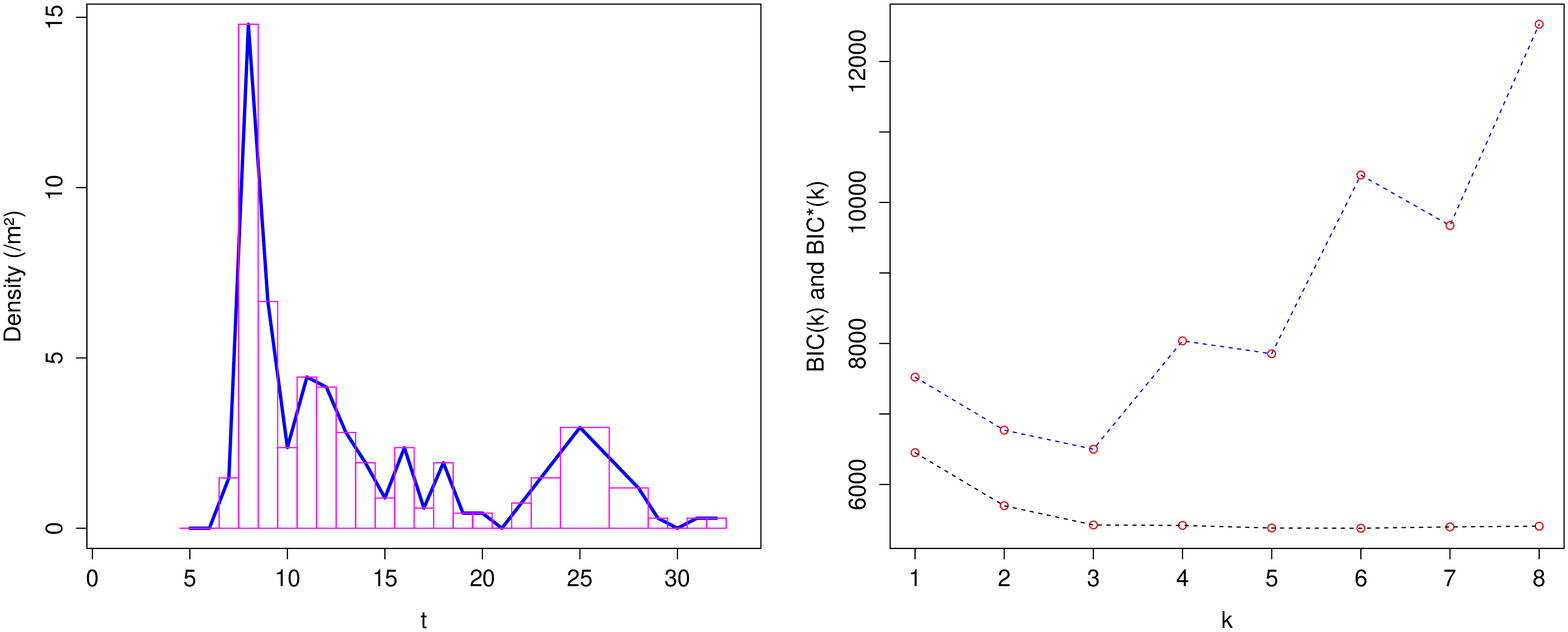}
\caption{\scriptsize{Example of flowering curve with its rectangulation and evolution of the associated BIC (in black) and $\textnormal{BIC}^{*}$ (in blue) from GMM estimation.}}
\label{FigExBICCurves4}
\end{figure}

\noindent As for the usual BIC, our choice relies on
$$
\wt{k}^{\,*} ~ = ~ \arg \min_{1\, \leq\, k\, \leq\, k_{m}} \textnormal{BIC}^{\,*}(k)
$$
for an arbitrary upper bound $k_{m}$. The estimation of $c$ is not difficult, it only has to ensure that $\wt{c}^{\,*} + \textnormal{BIC}(k) > 0$ and one can choose for example
$$
\wt{c}^{\,*} ~ = ~ \max_{1\, \leq\, k\, \leq\, k_{m}} \big\vert\, \textnormal{BIC}(k)\, \big\vert ~ -\min_{1\, \leq\, k\, \leq\, k_{m}} \textnormal{BIC}(k).
$$
To evaluate $\wt{\alpha}^{\,*}$, we make $\alpha$ vary on a grid and experiments are conducted on the learning set. Note that $c$ and $\alpha$ depends on $i$, meaning that each curve has its own parameters to be estimated in the criterion. From a practical point of view as we will see in the next section, $\wt{\alpha}^{\,*}$ only changes from a class of curves to another. On Figures \ref{FigExBICCurves1}--\ref{FigExBICCurves4} are represented four examples of flowering curves having different blooming behaviors supplied with their associated $g_{i}$ step functions. On the right, the evolution of BIC and $\textnormal{BIC}^{\,*}$ for $k_{m}=8$ is also given. As one can see on the graphs, $\textnormal{BIC}^{\,*}$ plays a moderation role and suggests to select $\wt{k}^{\,*} = 1,2, 3$ and 3 respectively, whereas BIC suggests to choose the quite unrealistic values $\wt{k} = 3, 8, 8$ and 8 respectively, for the aforementioned reasons. In fact as the reader will observe, for some curves (as the one of Figure \ref{FigExBICCurves2}), BIC suggests to select the maximal number of components whereas the common sense would have been to choose the same value of $k$ using BIC or BIC$^{*}$. However as we deal with hundreds of curves in each dataset, it is essential for us that an algorithmic procedure enables to select $k$, with no \textit{human} intervention.

\newpage

\newpage

\begin{figure}[H]
\centering
\includegraphics[width=16cm]{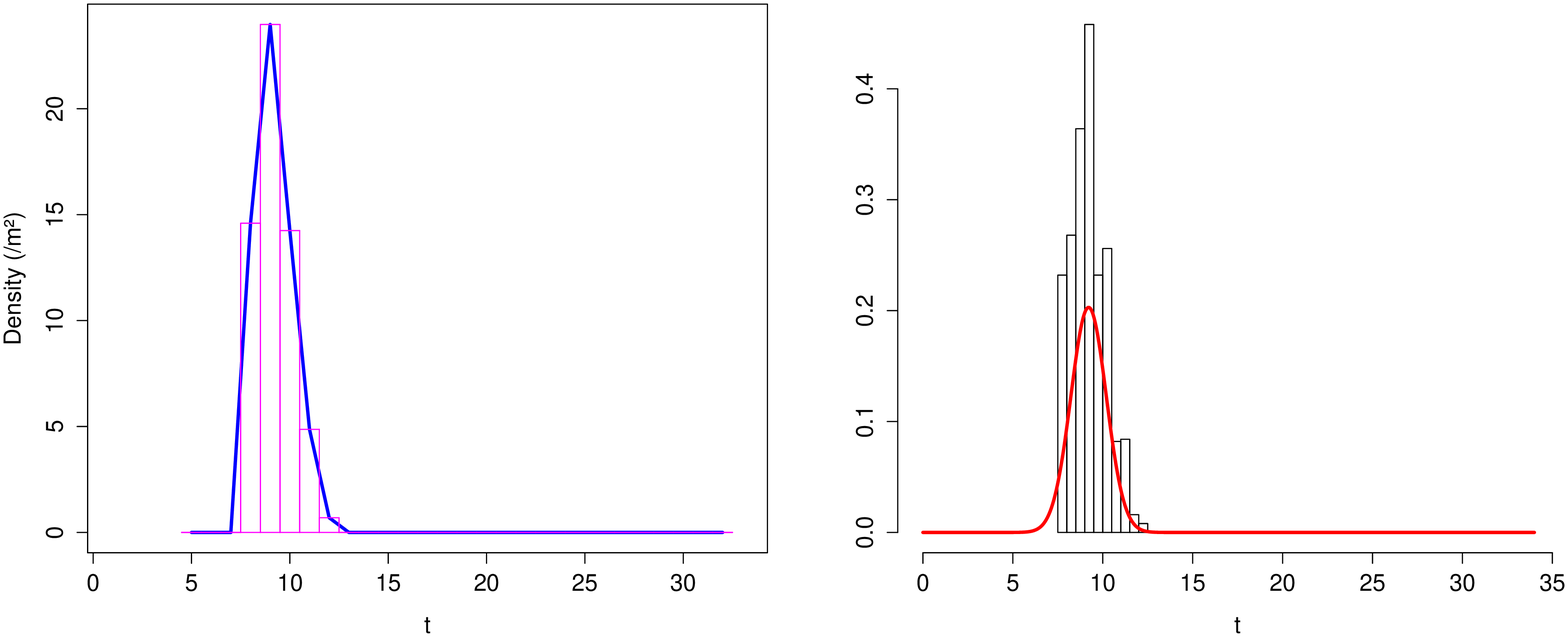}
\caption{\scriptsize{Example of the GMM algorithm running on a flowering curve, for $\wt{k}^{\,*} = 1$ selected cluster. On the right, the coloured curve is the estimated Gaussian component and the histogram of the generated sample is superimposed.}}
\label{FigExGMMCurves1}
\end{figure}

\begin{figure}[H]
\centering
\includegraphics[width=16cm]{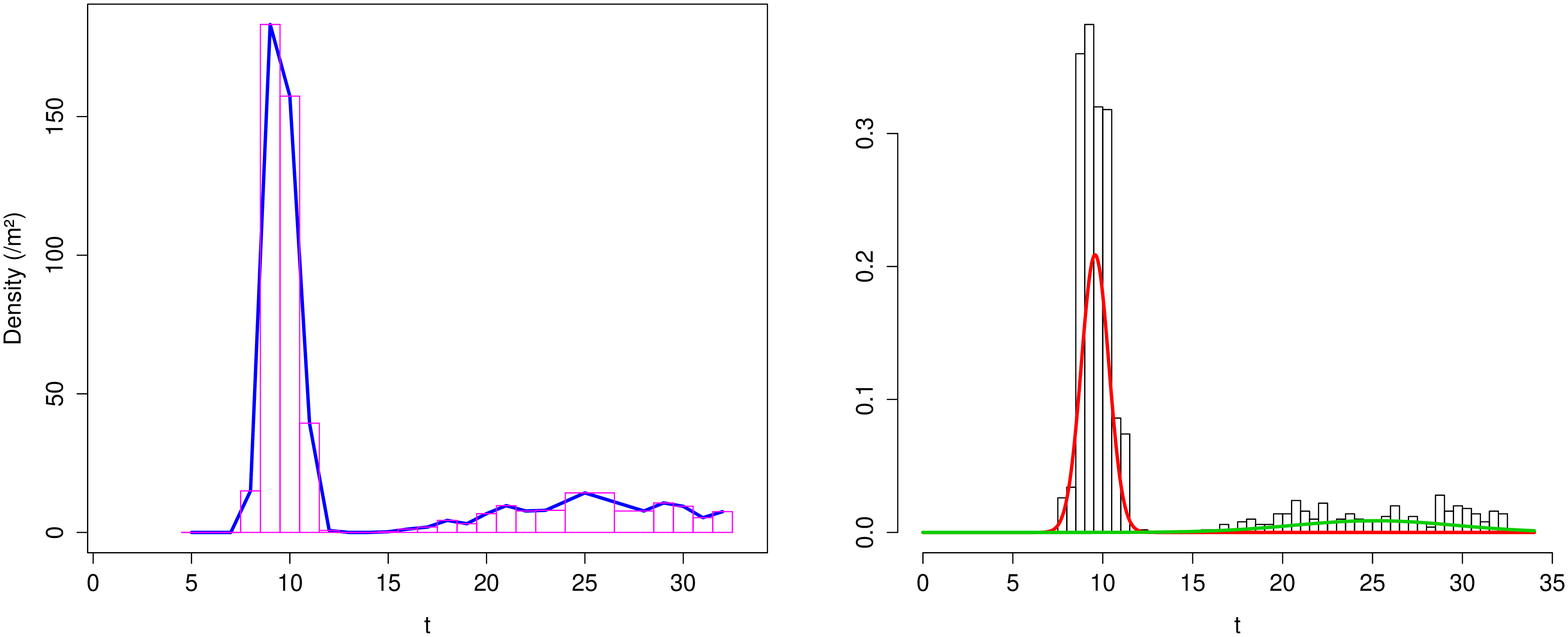}
\caption{\scriptsize{Example of the GMM algorithm running on a flowering curve, for $\wt{k}^{\,*} = 2$ selected clusters. On the right, the coloured curves are the estimated Gaussian components and the histogram of the generated sample is superimposed.}}
\label{FigExGMMCurves2}
\end{figure}

\begin{figure}[H]
\centering
\includegraphics[width=16cm]{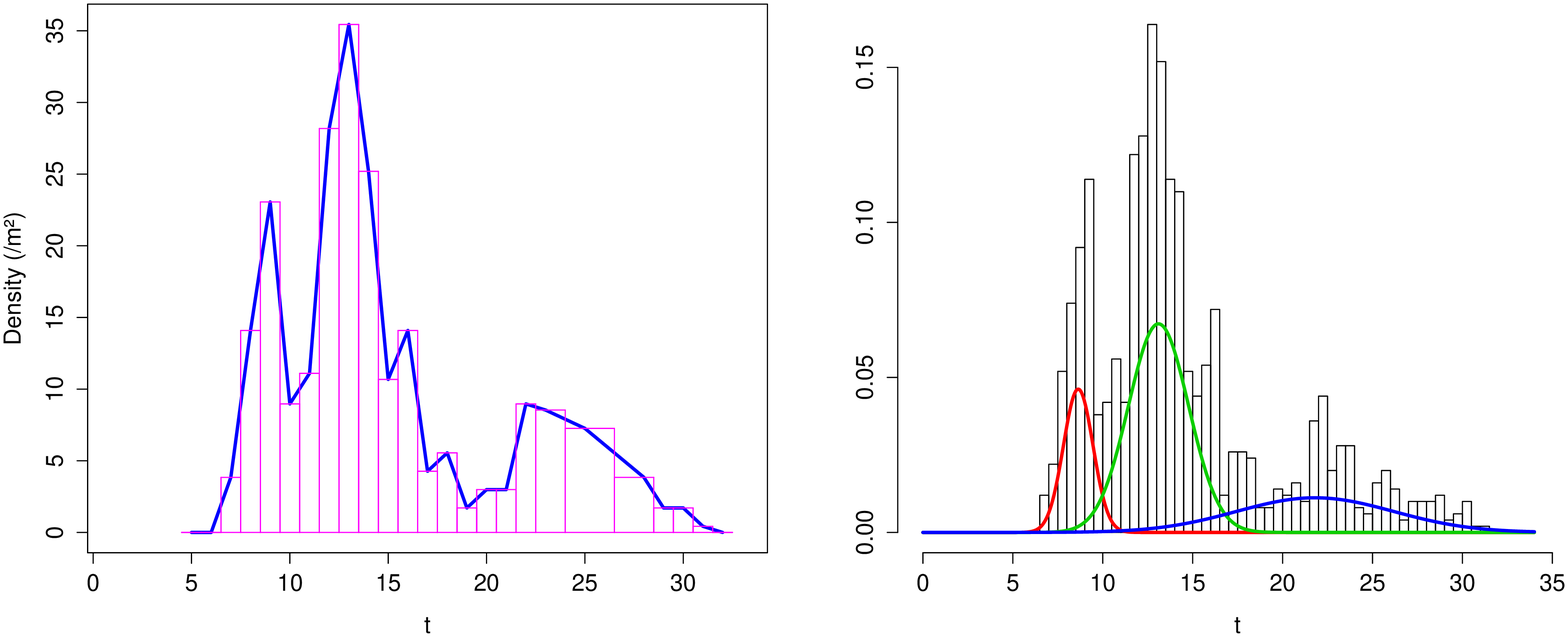}
\caption{\scriptsize{Example of the GMM algorithm running on a flowering curve, for $\wt{k}^{\,*} = 3$ selected clusters.  On the right, the coloured curves are the estimated Gaussian components and the histogram of the generated sample is superimposed.}}
\label{FigExGMMCurves3}
\end{figure}

\begin{figure}[H]
\centering
\includegraphics[width=16cm]{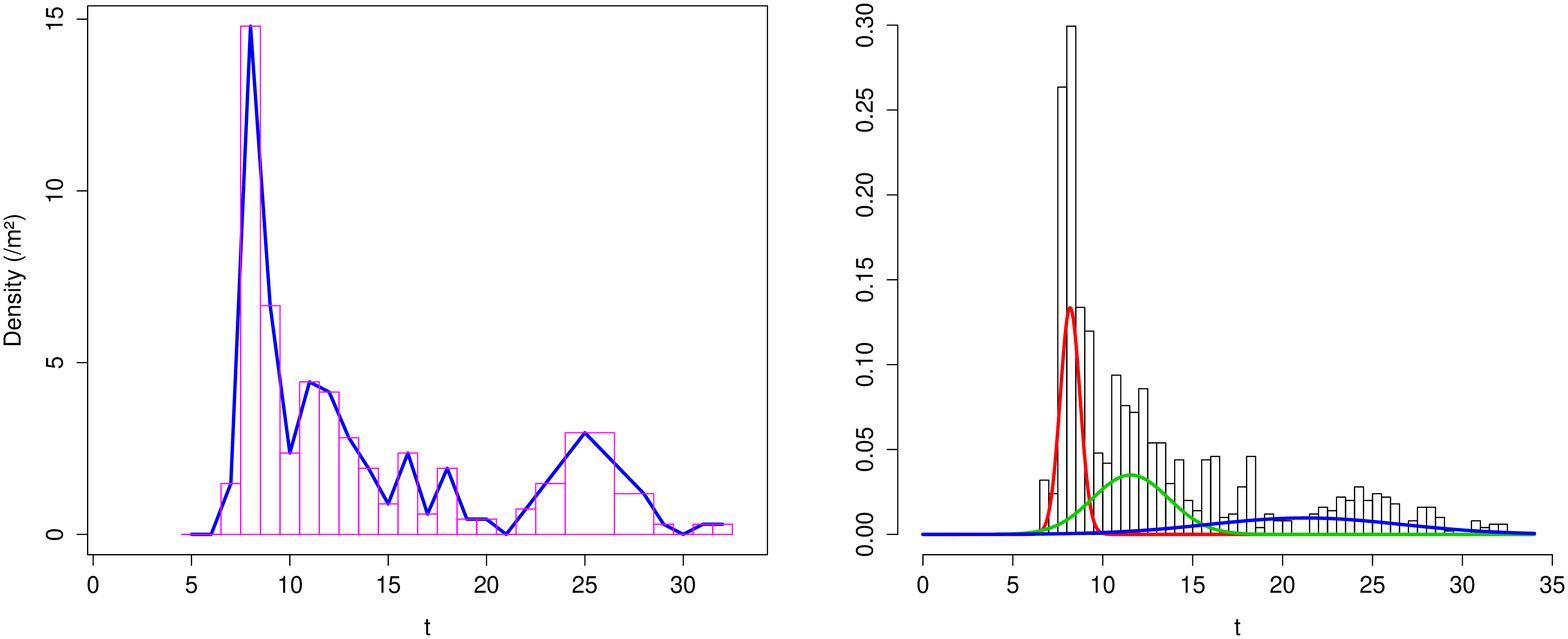}
\caption{\scriptsize{Example of the GMM algorithm running on a flowering curve, for $\wt{k}^{\,*} = 3$ selected clusters.  On the right, the coloured curves are the estimated Gaussian components and the histogram of the generated sample is superimposed.}}
\label{FigExGMMCurves4}
\end{figure}

\noindent The estimation of $c$ was automatically conducted and values $\wt{\alpha}^{\,*} = 2, 2.5, 2.5$ and $2.5$ were chosen for $\alpha$. The results of the GMM algorithm on these curves for $\wt{k}^{\,*}$ clusters are given on Figure \ref{FigExGMMCurves1}--\ref{FigExGMMCurves4} together with the histogram of the generated samples $(\wt{y}_{i,1}, \hdots, \wt{y}_{i,n_{s}})$ with $n_{s}=1000$.

\newpage

\newpage

\section{Indicators for PCA and Classification}
\label{SecClassif}

The dataset of flowering curves is very heterogeneous, this is the main motivation to find a set of indicators allowing to classify each rosebush. As for GMM, we also need to shortly summarize the main principles of the \textit{$k$-means for longitudinal data} algorithm \cite{GenoliniFalissard10} -- shortened from now on \textit{KML}. The objective is first to subdivide the dataset into classes according to the reblooming behavior of each curve, and then to provide a representative mean curve for each cluster of a subclassification.

\subsection{The $k$-means for longitudinal data}
\label{SecKML}

We consider a set of $n$ random vectors of $\dR^{d}$ defined as $Y_{i} = (Y_{i,1}, \hdots, Y_{i,d})$, for all $1 \leq i \leq n$. The KML algorithm merely works like an usual $k$-means algorithm on a set of paths $(y_{1}, \hdots, y_{n})$ that may be thought as $n$ time-related trajectories of length $d$. After convergence of the classification algorithm in $k$ clusters, denote by
$$
B_{k} = \frac{1}{n} \sum_{j=1}^{k} \wt{n}_{j}\, (\bar{y}_{j} - \bar{y})(\bar{y}_{j} - \bar{y})^{\prime}
$$
where $\wt{n}_{j}$ is the size of cluster $j$, $\bar{y}_{j}$ the corresponding mean trajectory and $\bar{y}$ the mean trajectory of the whole set. Define also
$$
W_{k} = \frac{1}{n} \sum_{j=1}^{k} \sum_{i=1}^{\wt{n}_{j}} (y_{i}^{j} - \bar{y}_{j})(y_{i}^{j} - \bar{y}_{j})^{\prime}
$$
where $y_{i}^{j}$ is the $i$--th path in cluster $j$. The between-variance matrix $B_{k}$ can be seen as an estimator of $\dV(\dE[Y \vert Z])$, where $Y$ stands for a representative random vector of an independent population and $Z$ is the latent classification random variable associated with $Y$, having $k$ modalities. Similarly, the within-variance matrix $W_{k}$ is an estimator of $\dE[\dV(Y \vert Z)]$. The number of clusters is selected by maximization of the Calinski--Harabasz criterion \cite{CalinskiHarabasz74} given by
$$
\textnormal{CH}(k) = \frac{(n-k)\, \textnormal{tr}(B_{k})}{(k-1)\, \textnormal{tr}(W_{k})}
$$
where $B_{k}$ and $W_{k}$ are estimated for $k$ clusters. The CH selection is then
$$
\wt{k} ~ = ~ \arg \max_{2\, \leq\, k\, \leq\, k_{m}} \textnormal{CH}(k)
$$
for an arbitrary upper bound $k_{m}$. The centroids of the $\wt{k}$ clusters will be seen as the representative curves of each class. To run KML algorithm, we will use the \texttt{R} library \texttt{kml} \cite{GenoliniAlacoqueSentenacArnaud15}. The opportunity to deal with temporal missing values was our main motivation to make use of KML instead of standard $k$--means for random vectors. Indeed as mentioned in the introduction, a non-negligible amount of data is missing in our curves, and a consistent completion was essential. In Figures \ref{FigKML1}--\ref{FigKML2} below, an example of KML is represented on simulated data with $n=35$ ($n_1 = 20$, $n_2=15$), $d=20$ and two different patterns with an additive noise, constrained to stay nonnegative. On the first figure, the patterns (chosen to look like flowering curves) are given, the generated curves together with their centroids found by the algorithm appear alongside. On the second figure, the evolution of the size of each cluster and the evolution of the CH criterion associated are given, for $k_{m}=6$. According to the CH criterion, $\wt{k} = 2$ is obviously selected.

\begin{figure}[H]
\centering
\includegraphics[width=16cm]{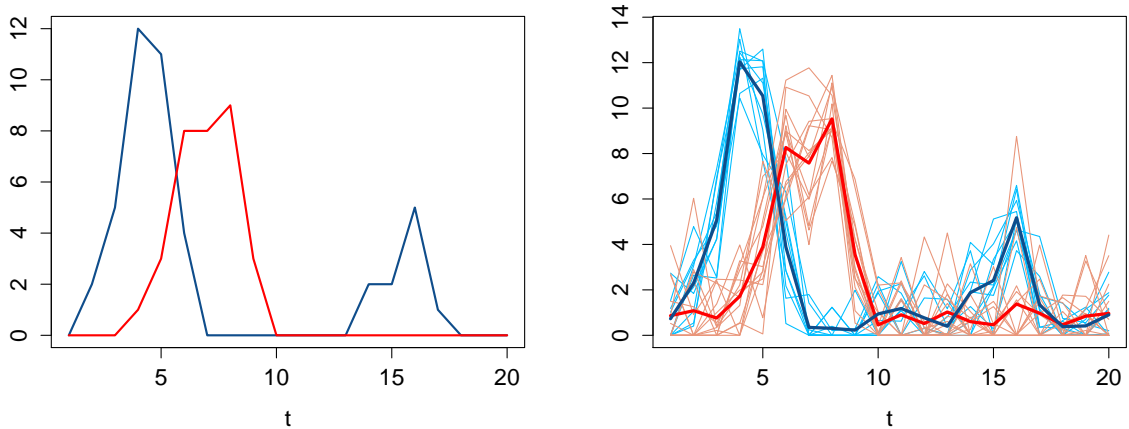}
\caption{\scriptsize{Example of KML running on a set of $n_1 = 20$ and $n_2 = 15$ simulated trajectories according to the patterns in the left, for $k=2$ and an additive noise. The centroids are highlighted.}}
\label{FigKML1}
\end{figure}

\newpage

\begin{figure}[H]
\centering
\includegraphics[width=16cm]{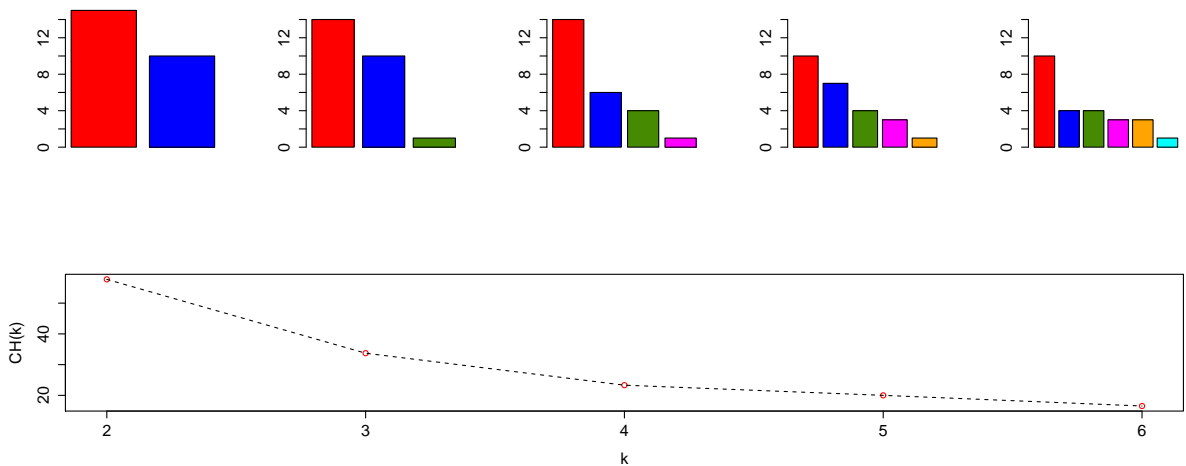}
\caption{\scriptsize{Evolution of the clusters size on the barplots and evolution of the CH criterion related to the above example, for $k_{m}=6$.}}
\label{FigKML2}
\end{figure}

\subsection{A set of indicators and a reblooming classification}

\begin{figure}[H]
\centering
\includegraphics[width=16cm]{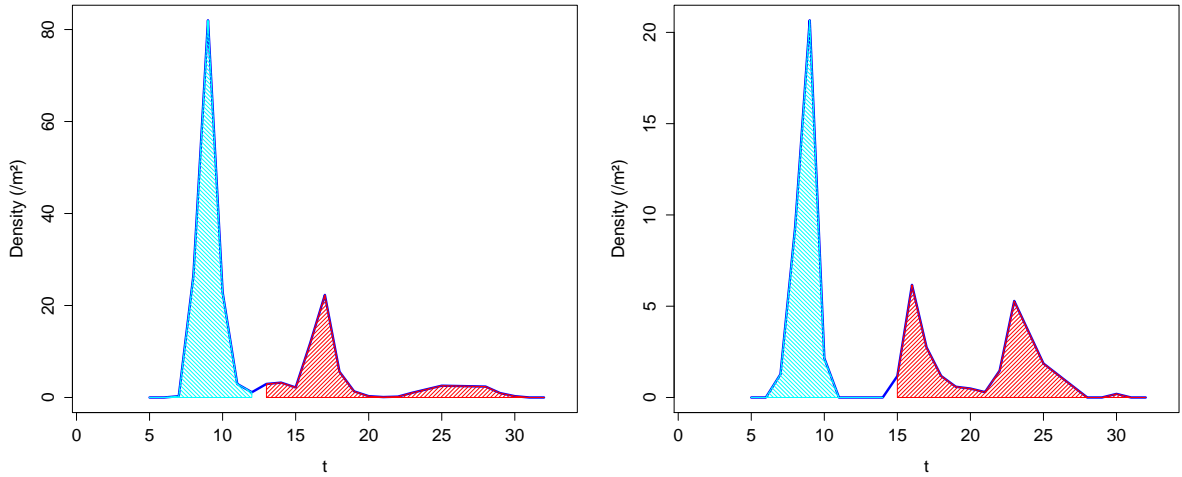}
\caption{\scriptsize{Examples of separation of the main peak area and of the reblooming area for two flowering curves. The first significative flowering is tinted in blue and the reblooming area detected by our algorithm is tinted in red.}}
\label{FigExAreas1}
\end{figure}

\begin{figure}[H]
\centering
\includegraphics[width=16cm]{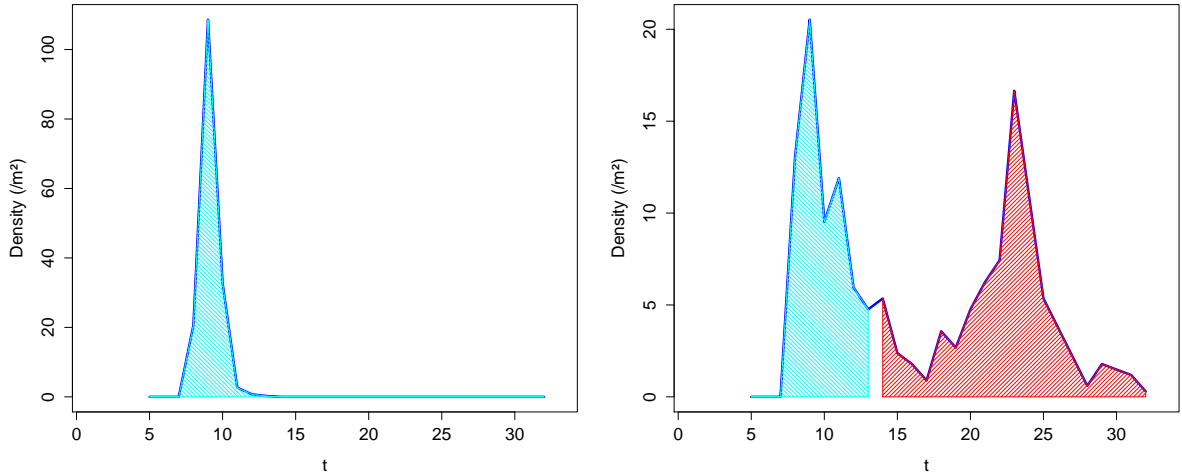}
\caption{\scriptsize{Examples of separation of the main peak area and of the reblooming area for two flowering curves. The first significative flowering is tinted in blue and the reblooming area detected by our algorithm is tinted in red.}}
\label{FigExAreas2}
\end{figure}

Let us denote by $t_{i, R}$ the first instant of significative reblooming of the $i$--th rosebush, that is the first time where the density exceeds a given threshold after the end of its first flowering. This threshold is calculated as a percentage (25\% in our examples) of the peak value during the first significative flowering, itself detected using a similar algorithm. Consider the areas
\begin{equation}
\label{Aires}
P_{i} = \int_{t_{i,0}}^{t_{i, R}} g_{i}(x)\, \dd x \hsp \text{and} \hsp R_{i} = \int_{t_{i, R}}^{t_{i, d+1}} g_{i}(x)\, \dd x
\end{equation}
with the convention that $t_{i, R} = t_{i, d+1}$ if the $i$--th rosebush does not have a significative reblooming (we recall that $d$ is the observation vector size), and where $g_{i}$ is the step function given in \eqref{StepFunc}. $P_{i}$ stands for the area under the first significative flowering and $R_{i}$ for the area under the whole reblooming period, possibly zero. On Figures \ref{FigExAreas1} and \ref{FigExAreas2}, we present four examples of such detection, leading to $t_{i,R} = 13$ (top left), $15$ (top right), $33$ (bottom left) and 14 (bottom right). The areas $P_{i}$ and $R_{i}$ are tinted, respectively in blue and red (note that, strictly speaking, it is only an approximation of the areas which is coloured, we have not represented the rectangulations on these graphs, which are the effective tools used to compute the areas).

\newpage

\noindent Now we consider the \textit{reblooming magnitude} that we define as
\begin{equation}
\label{AmpRem}
M_{i} = \frac{R_{i}}{P_{i}+R_{i}}.
\end{equation}

\noindent This indicator is going to play a primordial role in our classification. Literally, we have built the OF/R$_1$/R$_2$ classes by using a $k$-means algorithm on the first PCA plane (see Figure \ref{FigACPIndiv} later) according to the indicators that we are going to describe. OF stands for \textit{once-flowering}, R$_1$ and R$_2$ for \textit{weak} and \textit{strong reblooming}. The numeric indicators that we have decided to retain to form a basis of comparison between the whole curves are given below. They come either from the plant features or from the statistical analysis. We use the notations $\wt{\pi}_{\textnormal{max}} = \max(\wt{\pi}_1, \hdots, \wt{\pi}_{\wt{k}^{*}})$, $\wt{\mu}_{\textnormal{min}} = \min(\wt{\mu}_1, \hdots, \wt{\mu}_{\wt{k}^{*}})$ and $\wt{\sigma}^{\, 2}_{\textnormal{min}} = \min(\wt{\sigma}^{\, 2}_1, \hdots, \wt{\sigma}^{\, 2}_{\wt{k}^{*}})$, corresponding to the GMM estimations. Among all sets of indicators on which a PCA has been conducted, the following one has given the most meaningful results.
\begin{itemize}
\item \texttt{Nb.Clust}: the number $\wt{k}^{\,*}$ of clusters selected by our BIC$^{*}$ criterion in the GMM applied on the curve.
\smallskip
\item \texttt{LMax.P}: $\ln(\wt{\pi}_{\textnormal{max}})$, the log-weight affected to the main flowering.
\smallskip
\item \texttt{Rat.P}: $(1-\wt{\pi}_{\textnormal{max}})/\wt{\pi}_{\textnormal{max}}$, the ratio between the weight affected to the reblooming area and the weight of the main flowering.
\smallskip
\item \texttt{Min.M}: $\wt{\mu}_{\textnormal{min}}$, the lowest estimated mean.
\smallskip
\item \texttt{DifMax.M}: $d_{\wt{k}^{\,*}}$ as it is defined in \eqref{DifMin} for $\wt{k}^{\,*} \geq 2$, the highest difference between the estimated means. By convention, difference between the last and first instants of measure for $\wt{k}^{\,*}=1$.
\smallskip
\item \texttt{Min.V}: $\wt{\sigma}^{\, 2}_{\textnormal{min}}$, the lowest estimated variance.
\smallskip
\item \texttt{Area.Peak.Inflo}: the area under the main flowering, defined as $P_{i}$ in \eqref{Aires}, divided by the mean number of flowers within an inflorescence.
\smallskip
\item \texttt{Reb.Mag}: the reblooming magnitude defined as $M_{i}$ in \eqref{AmpRem}.
\smallskip
\item \texttt{Max.Peak}: the highest value of the first significative flowering.
\smallskip
\item \texttt{Max.Reb}: the highest value of the reblooming period.
\smallskip
\item \texttt{First.Reb}: the first instant of the reblooming period called $t_{i, R}$ in \eqref{Aires}, last instant of measure if there is no significative reblooming.
\smallskip
\item \texttt{Rat.Peak}: the ratio between \texttt{Max.Reb} and \texttt{Max.Peak}.
\smallskip
\item \texttt{First.Flo}: the first instant of flowering.
\smallskip
\item \texttt{Cont.Flo}: an indicator in $\{ 0,1 \}$ characteristic of plants having a continuous significative flowering. In the framework of this study, a continuous flowering is a feature of a curve taking most of the time ($> 80\%$) positive values.
\end{itemize}

\newpage

\begin{figure}[H]
\centering
\includegraphics[width=16cm]{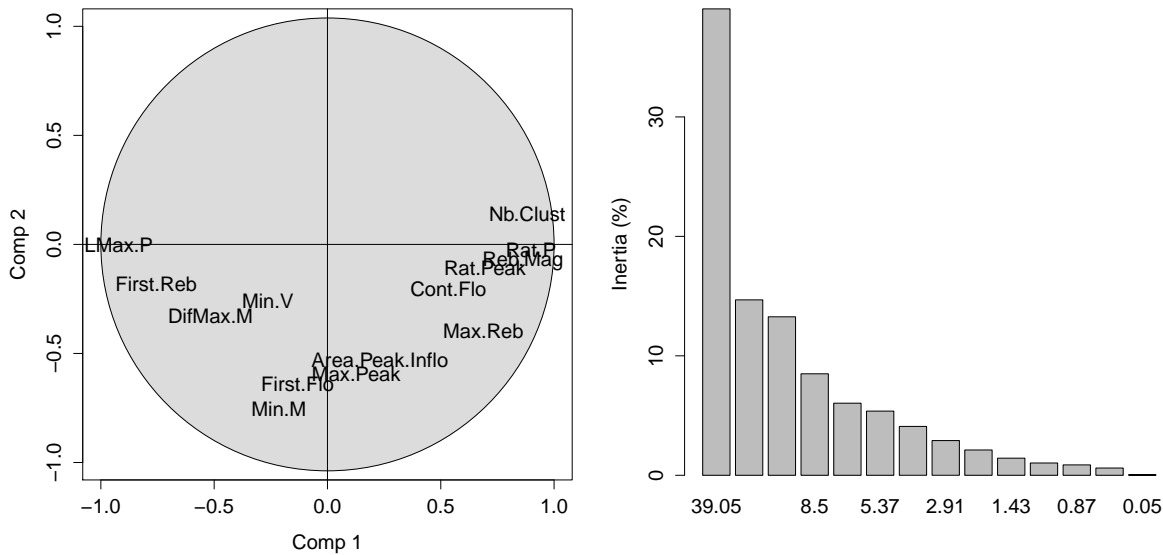}
\caption{\scriptsize{Projection of the indicators on the main factorial plane of the PCA on the left, and barplot of the percentage of inertia associated with each axis on the right.}}
\label{FigACPVarInert}
\end{figure}

\noindent The PCA analysis reveals a dominant eigenvalue expressed on the first axis and another eigenvalue far less significative, expressed on the second axis. In the context of the study, we will content ourselves with this couple of eigenvalues. On Figure \ref{FigACPVarInert}, all our indicators have been projected on the main factorial plane and we make the following observations: the reblooming and the precocity seem to form the main orthogonal features of a rosebush, in order of importance. To be complete, let us precise that more indicators have been initially tested (\texttt{LMin.P}, \texttt{Max.M}, \texttt{Max.V}, \texttt{Moy.M} and \texttt{Moy.V} (mean values), \texttt{DifMax.M}, \texttt{Inflo} (intensity of inflorescence), \texttt{Area.Peak}, etc.) With equivalent results (more than 0.95 of empirical correlation between \texttt{Area.Peak} and \texttt{Max.Peak}, for example), the smallest set has been preferred, for purposes of parsimony. One can already notice that a few curves are considered as OF whereas their reblooming magnitudes are non-zero: according to us, this is due either to artefacts for biological or environmental reasons, or imprecisions in gathering data.

\subsubsection{The reblooming indicators}

Six indicators are strongly correlated to describe the reblooming behavior of a rosebush: \texttt{Nb.Clust}, \texttt{Reb.Mag}, \texttt{Rat.P}, \texttt{Rat.Peak}, \texttt{Cont.Flo} and \texttt{LMax.P}, the reasons being obvious for most of them. According to our model, the reblooming deepens when $\wt{\pi}_{\textnormal{max}}$ decreases, this explains the fact that $\texttt{Rat.P}$ is positively and \texttt{LMax.P} negatively correlated with the reblooming behavior. \texttt{First.Reb} and \texttt{DifMax.M} play a negative role: \texttt{DifMax.M} is affected by the large values of once-flowering curves whereas \texttt{First.Reb} increases when reblooming starts later. The positive correlation of \texttt{Max.Reb} with the reblooming phenomenon seems also quite obvious.

\newpage

\subsubsection{The precocity indicators}

Four indicators are related to the precocity of the rosebush: \texttt{First.Flo}, \texttt{Min.M}, \texttt{Area.Peak.Inflo} and \texttt{Max.Peak}. Some of them have a trivial explanation, whereas it is quite unexpected for us to observe that the main flowering seems positively related to the precocity: a rosebush having a later first flowering produces on average a more abundant first flowering. A more favourable weather might be a trail to explain it.

\bigskip

\noindent Finally, the last indicator \texttt{Min.V} does not appear to play any role in the reblooming or the precocity of the rosebush. To sum up, the reblooming features are clearly described by the classification OF/R$_1$/R$_2$. In the following section, we aim to look for a subclassification of the curves based on the precocity indicators. As we will see, this is unfortunately far less convincing. There is an effect of the second axis on the subclasses but it is not as perceptible as the effect of the first axis on the classes.

\begin{figure}[H]
\centering
\includegraphics[width=15cm]{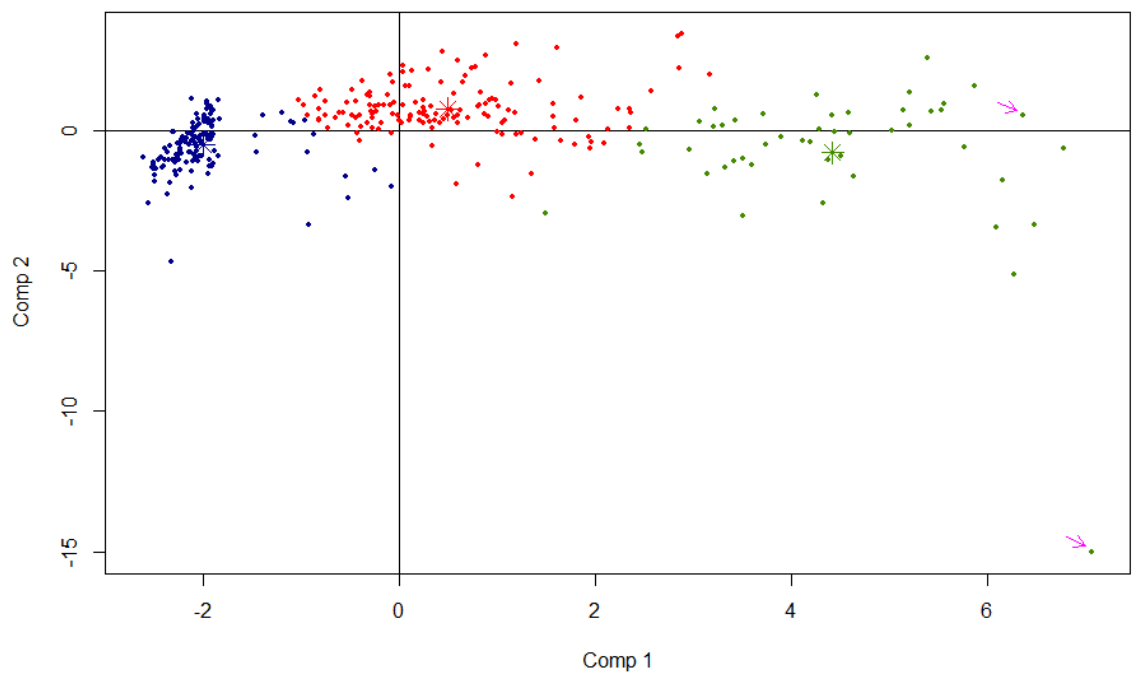}~\bigskip
\caption{\scriptsize{Projection of the individuals on the main factorial plane of the PCA. From a $k$-means classification, OF are coloured in blue, R$_1$ in red and R$_2$ in green. The checked points correspond to the flowering curves of Figure \ref{FigExAreasOpp} and the centroids are indicated.}}
\label{FigACPIndiv}
\end{figure}

\begin{figure}[H]
\centering
\includegraphics[width=16cm]{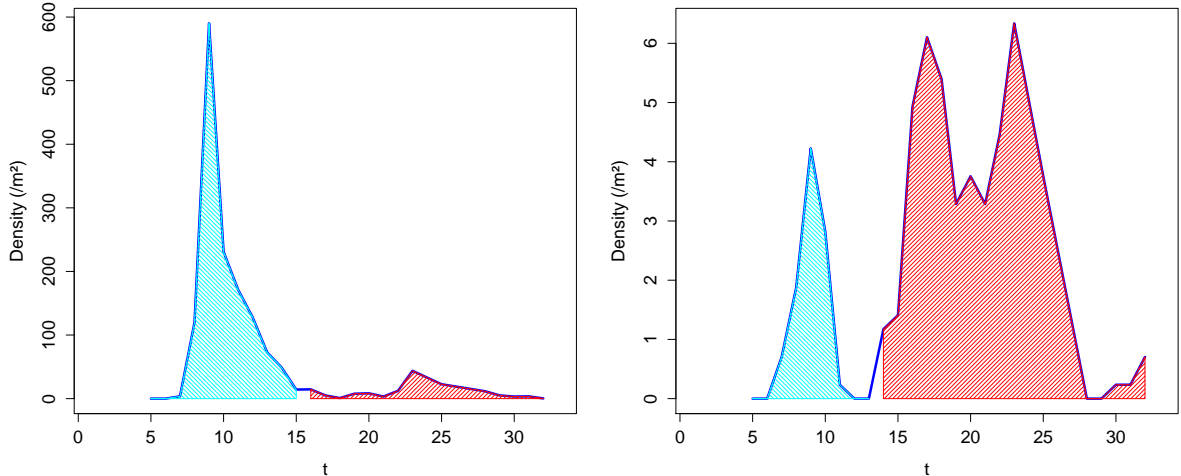}
\caption{\scriptsize{Examples of the peak/reblooming areas associated with the flowering curves checked in Figure \ref{FigACPIndiv}. The first one has a massive flowering despite a low reblooming magnitude whereas the second one has the highest reblooming magnitude of the dataset.}}
\label{FigExAreasOpp}
\end{figure}

\noindent The fuzzy boundary between green (R$_2$) and red (R$_1$) individuals is not problematic in our framework since R$_1$ and R$_2$ are only separated on a \textit{quantitative} basis. The presence of blue (OF) individuals in the red (R$_1$) area, however separated on a \textit{qualitative} basis, seems somehow more annoying. They correspond to the aforementioned curves having a non-zero reblooming but weak enough to be considered as artefact. For R$_2$, the lack of concentration of the individuals in the factorial plane is justified by the strong heterogeneity of the reblooming curves, despite the numerous descriptive indicators. By way of example, we have represented on Figure \ref{FigExAreasOpp} the flowering curves checked on Figure \ref{FigACPIndiv}.

\noindent Both of them are R$_2$ curves. The first one has a quite low reblooming magnitude ($\approx 0.14$) but the flowering is so abundant that the peak indicators, though weakly correlated with the first axis, project the rosebush to the bottom-right corner of the factorial plane. The second one has the highest reblooming magnitude of the dataset ($\approx 0.86$) and is accordingly also located at the right of the plane. These examples show the huge heterogeneity of the reblooming classes. This prior classification leads to consider a substantial part of the dataset as once-flowering rosebushes whereas reblooming classes are dominated by weakly reblooming plants. The sizes are the following: on $n=329$ exploitable flowering curves, $n_0=152$ are OF, $n_1=127$ are R$_1$ and $n_2=50$ are $R_2$ (see Figure \ref{FigPropOFR} below for a recap chart of the subclasses).

\subsection{A subclassification of the curves using KML}

\begin{figure}[H]
\centering
\includegraphics[width=11.2cm]{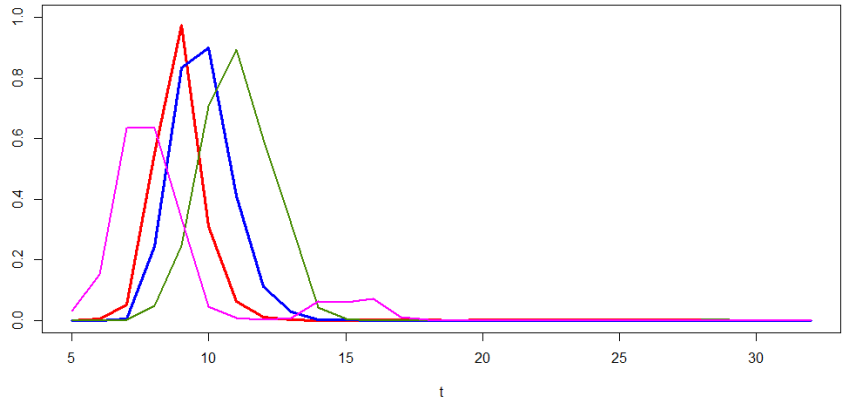}~\bigskip
\caption{\scriptsize{Representative curves of the $\wt{k}=4$ subclasses of the OF standardized dataset, from the KML algorithm.}}
\label{FigClustMoyOF}
\end{figure}

\begin{figure}[H]
\centering
\includegraphics[width=11.2cm]{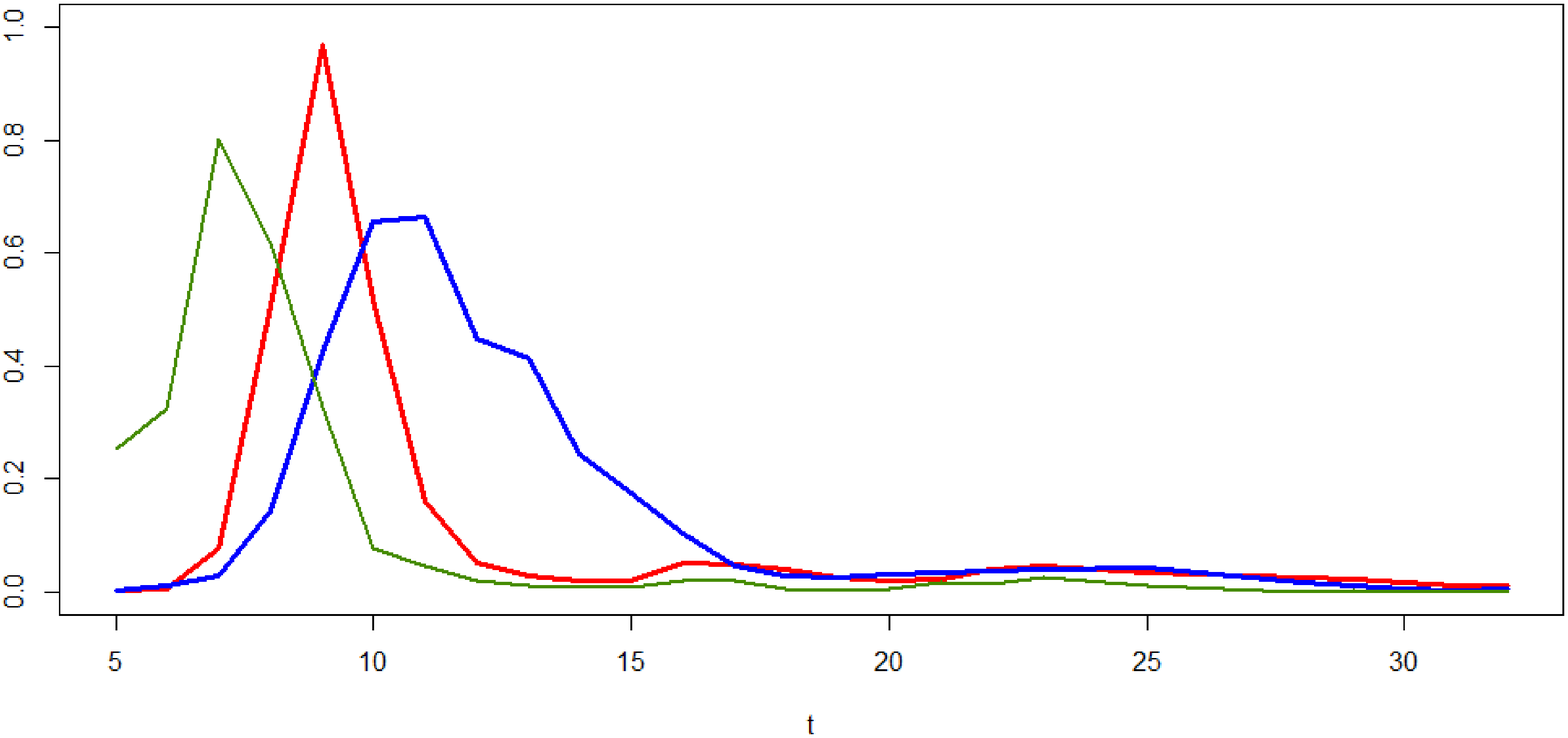}~\bigskip
\caption{\scriptsize{Representative curves of the $\wt{k}=3$ subclasses of the R$_1$ standardized dataset, from the KML algorithm.}}
\label{FigClustMoyR1}
\end{figure}

\begin{figure}[H]
\centering
\includegraphics[width=11.2cm]{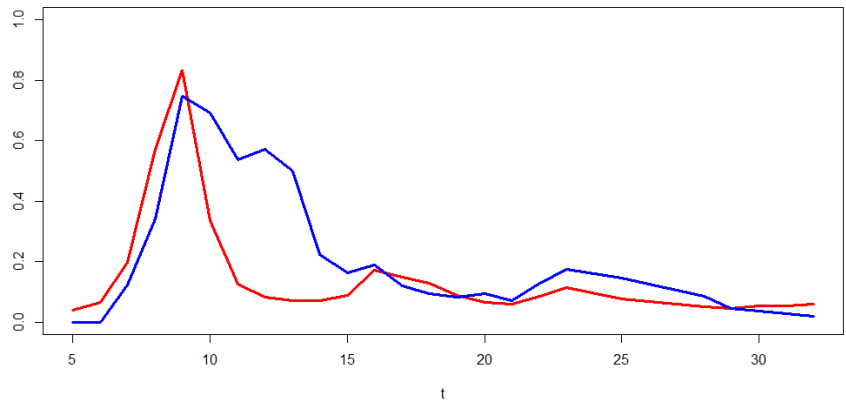}~\bigskip
\caption{\scriptsize{Representative curves of the $\wt{k}=2$ subclasses of the R$_2$ standardized dataset, from the KML algorithm.}}
\label{FigClustMoyR2}
\end{figure}

\noindent The next step is to run the KML algorithm in each class of curves (OF/R$_1$/R$_2$), to highlight similar behaviors. We start by standardizing the dataset so as to restrict any scale effect. For all $1 \leq \ell \leq d$, let
$$
z_{i,\ell} = \frac{y_{i,\ell}}{\Vert y_{i} \Vert_{\infty}}
$$
where $y_{i} = (y_{i,1}, \hdots, y_{i,d})$ is the observed path associated with the $i$--th curve of the dataset, and $\Vert \cdot \Vert_{\infty}$ is the usual infinity norm of $\dR^{d}$. On the basis of the CH criterion (see Section \ref{SecKML}) and looking for reasonable sizes in each subclass, $\wt{k} = 4, 3$ and 2 clusters are selected for OF, R$_1$ and R$_2$ respectively. The representative curves of each subclass are given on Figure \ref{FigClustMoyOF}--\ref{FigClustMoyR2}.

\newpage

\newpage

\begin{figure}[H]
\centering
\includegraphics[width=10cm]{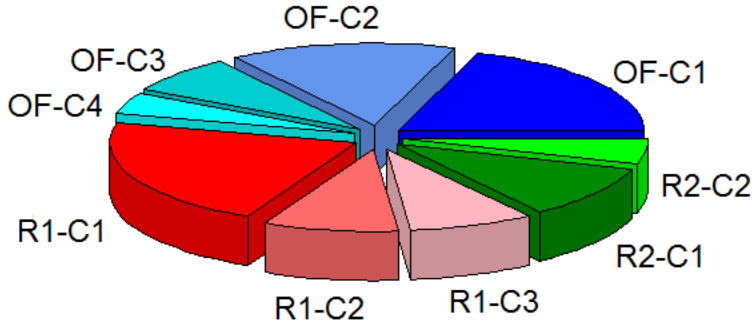}~\bigskip
\caption{\scriptsize{Descriptive statistics: pie chart associated with the proportions of the reblooming classes OF/R$_1$/R$_2$ in the dataset and with their KML subclasses.}}
\label{FigPropOFR}
\end{figure}

\noindent The proportions of the whole classification are represented thereafter in Figure \ref{FigPropOFR}. Among the $n_0=152$ curves in the OF class, $n_{01} = 65$ are $C_1$ (blue centroid), $n_{02} = 49$ are $C_2$ (red centroid), $n_{03} = 24$ are $C_3$ (green centroid) and $n_{04} = 14$ are $C_4$ (magenta centroid). Similarly, among the $n_1=127$ curves in the R$_1$ class, $n_{11} = 71$ are $C_1$ (blue centroid), $n_{12} = 29$ are $C_2$ (red centroid) and $n_{13} = 27$ are $C_3$ (green centroid). Finally, among the $n_2=50$ curves in the R$_2$ class, $n_{21} = 34$ are $C_1$ (blue centroid) and $n_{22} = 16$ are $C_2$ (red centroid).

\noindent  The main indicator of selection is clearly the reblooming magnitude, but the representative curves highlights the orthogonal indicator lying in the precocity of the first flowering peak. It is especially perceptible on the once-flowering (OF) and weakly reblooming (R$_1$) curves, whereas it seems to vanish on the highly reblooming (R$_2$) curves where the chaotic behavior yields difficulties to clearly identify the beginning of the flowering.


\section{Concluding Remarks}
\label{SecConclu}

According to the authors, this work is a further evidence of the well-established consistence of Gaussian mixture models with biological data. The ability of a Gaussian mixture to smooth the observations and to highlight hidden structures form a convincing answer to the main features of flowering curves. In addition, the \textit{waves} mechanism is not only a suitable statistical modelling, it also has a biological and genetic interpretation. The modified BIC$^{*}$ criterion we have suggested to use may be seen as an artificial correction of the potentially asymmetry of the flowering phenomenon, and somehow interpreted as a simplistic solution. Hence, non-Gaussian mixture models could be used for further investigations. The authors do not exclude the possibility that a peak of flowering may be better explained by an almost Gaussian but slightly asymmetric distribution. A deep statistical investigation of the symmetry in the OF dataset should lead to sufficient evidence, nevertheless the parameters estimation will pose some significant issues.

\smallskip

By comparison, the subclassification using the $k$-means longitudinal algorithm is far less meaningful. Not surprisingly, the precocity of the rosebushes is the main highlighted feature and, as we have seen in this work, a reasonable interpretation is only possible in the OF class. The authors have also considered the eventuality of a deformation model in which, for all $1 \leq \ell \leq d$,
$$
Y_{i,\ell} = a_{i}\, R_{i, \ell-t_{i}} + \veps_{i, \ell}
$$
where $(Y_{i,1}, \hdots, Y_{i,\ell})$ is the random flowering curve of the $i$--th rosebush, $a_{i}$ and $t_{i}$ are real parameters, $(R_{i,1}, \hdots, R_{i,\ell})$ is the representative curve of the associated KML subclass and $(\veps_{i,1}, \hdots, \veps_{i,\ell})$ is a white noise sequence. Thus, each flowering curve was seen as a linear deformation by centering/distension of its representative KML curve. Owing to the heterogeneity of the R$_1$/R$_2$ datasets, results obtained from a standard least squares approach were not satisfactory for our purposes. The estimations of $a_{i}$ and $t_{i}$ could have been relevant alternatives to \texttt{Max.Peak} and \texttt{First.Flo} in our PCA, but probably they would have led to the same kind of conclusions. Nevertheless, it seems that the two main orthogonal features characterizing a flowering curve are the reblooming phenomenon (number of clusters, magnitude, etc.) and the precocity (first instant of flowering, lowest estimated mean, etc.).

\smallskip

The fact that reblooming and precocity are found independent in this sample may have a genetical reason. Recurrent blooming is mainly controlled by a recessive locus that was recently identified as \textit{RoKSN} by Iwata \textit{et al.} \cite{Iwata12}. The date of the first flowering has a more complex genetic determinism with different loci controlling this trait, as shown by Kawamura \textit{et al.} \cite{Kawamura11} and Roman \textit{et al.} \cite{Roman15}. The QTL having the main effect was proposed to correspond to the gene \textit{RoFT} (see Otagaki \textit{et al.} \cite{Otagaki15}). The work of Spiller \textit{et al.} \cite{Spiller11} has established that \textit{RoKSN} and \textit{RoFT} map respectively to linkage groups 3 and 4. Due to the genetic independence between these genes, specific alleles for these genes have not been conjointly inherited, and therefore these phenotypic traits are not correlated. Previously, Kawamura \textit{et al.} \cite{Kawamura11} and Roman \textit{et al.} \cite{Roman15} have shown a weak correlation between recurrent blooming and the precocity by studying F$_1$ progenies. In these progenies, reblooming rosebushes flower earlier than once-flowering ones. This difference can be explained by a genetic linkage between the recurrent blooming locus and a precocity locus (genes involved in gibberellic acid are in the vicinity and are good candidates). In the current study, such association is not found because the genetic basis is larger (more than 300 individuals) than in F$_1$ progeny (two parents) and linkage disequilibrium between two linked genes is more likely to be reduced by the number of meioses during rose history. 

\smallskip

Environmental and genetic effects are not entirely explicable and separable due to the lack of repetitions of the measures. Indeed, a weather effect can not be neglected as flowering is highly controlled by environmental factors (see Andr\`es and Coupland \cite{AndresCoupland12} for a review). Likewise, in our probabilistic model all rosebushes are considered as independent. This hypothesis is questionable: we have every biological reasons to think that two plants growing in the same environmental conditions show some similitudes. Some rosebushes are also genetically close, which can lead to artificial associations between flowering parameters. Another direction for a future study lies in the consistence and the persistence of our conclusions on a temporal and spatial prolongation, and in the comparative flowering behavior of genetically close rosebushes. Such an extended study could also provide answers to some questions of great biological interest, like the relation between the intensity of the first flowering of a rosebush and its reblooming nature.

\smallskip

Since the work of Semeniuk \cite{Semeniuk71} in 1971 in \textit{Rosa wichurana} and recent molecular characterization by Iwata \textit{et al.} \cite{Iwata12} in 2012, reblooming is considered to be controlled by a recessive allele of a single gene (\textit{RoKSN}). Therefore, reblooming has until now been treated as a qualitative trait with as modalities `recurrent' and `non-recurrent' blooming roses (see de Vries and Dubois \cite{DeVriesDubois84}) or sometimes `once-flowering', `occasionally reblooming' and `continuous flowering' roses (see Iwata \textit{et al.} \cite{Iwata12}). The complexity of flowering curves and the analysis displayed in Figure \ref{FigACPIndiv} show that the reblooming trait should be treated as a semi-quantitative trait (once-flowering roses and roses with various reblooming intensities). The large sample size and its large diversity by comparison with previous studies probably explain the detection of the heterogeneity in flowering curves. Data produced and statistical tools developed in this work pave the way for the investigation of the environmental and genetic causes of this heterogeneity. This work is a preliminary step for finding new genes or new alleles controlling lower reblooming intensity than the one conferred by the Chinese copia allele of \textit{RoKSN}. As an example, \textit{Rosa damascena} `Four Seasons', with occasionnal reblooming, was cultivated in the Middle East then in Europe long before the introduction of the first Chinese roses at the end of eighteenth century. Genetic methods like association mapping (see Balding \cite{Balding06} for a review), aiming at finding correlations between genetic markers and phenotypic traits in germplasm, may allow to complete the knowledge of both reblooming genetic control and rose breeding history.

\smallskip

\noindent \textbf{Acknowledgments.} \textit{The acquisition of flowering curves was supported by the ``R\'egion Pays de la Loire'' in the framework of the FLORHIGE project, and by the department BAP of INRA in the framework of the SIFLOR project. We thank Th\'er\`ese and Raymond Loubert for providing access to their rose garden. We also thank Rachid Boumaza for his advices, Laure Ballerie for her work, Fabrice Foucher for his critical and pertinent view along the project, and the courageous people spending days and days making flowering measurements. Finally, we thank the Associate Editor and the two anonymous Reviewers for their suggestions and constructive comments which helped to improve substantially the paper.}

\newpage


\section*{Appendix. The Waves Mechanism of Flowering}
\label{AppWaves}
\renewcommand\thefigure{A.\arabic{figure}} 
\begin{figure}[H]
\centering
\includegraphics[width=4.9cm]{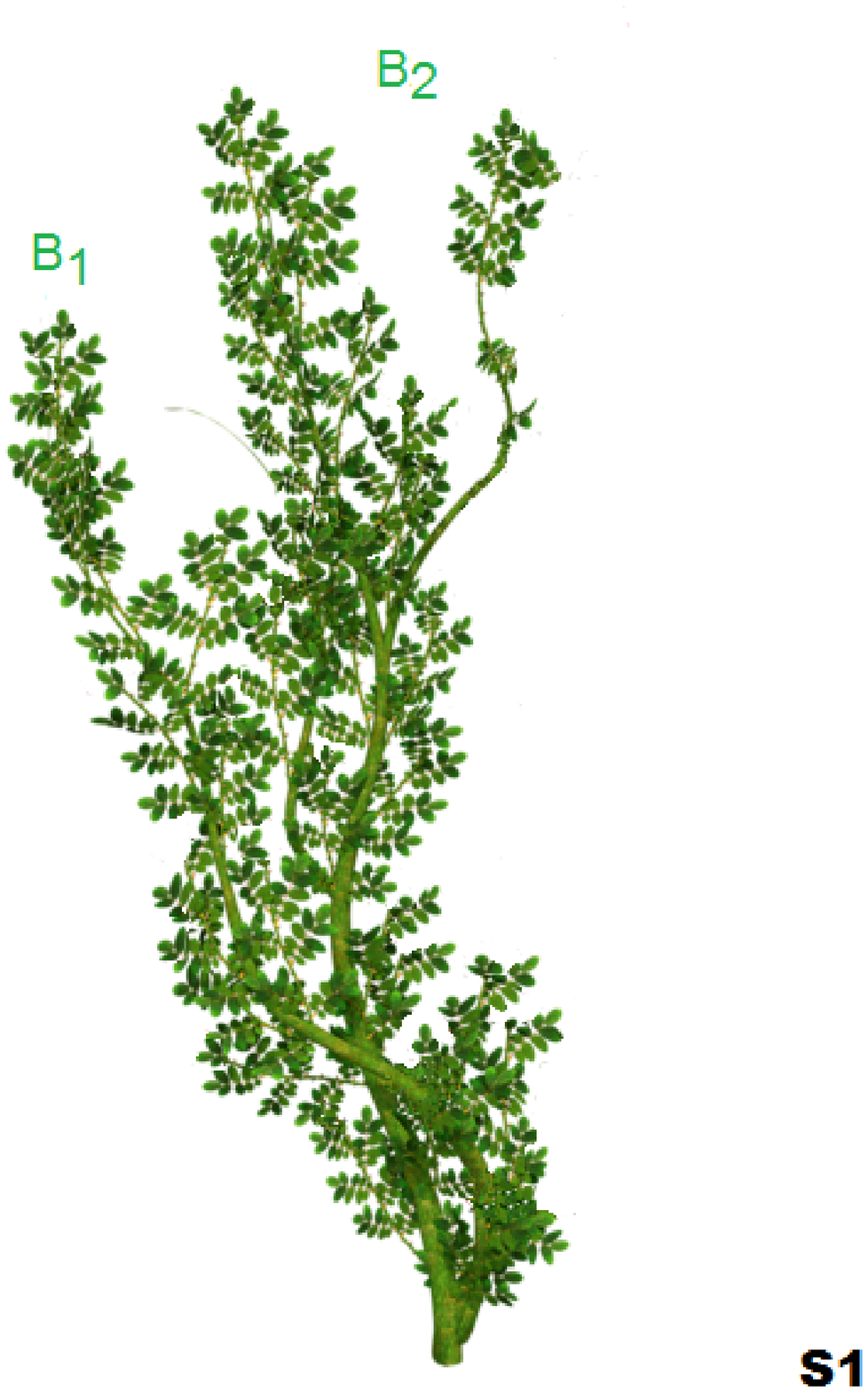} \includegraphics[width=4.9cm]{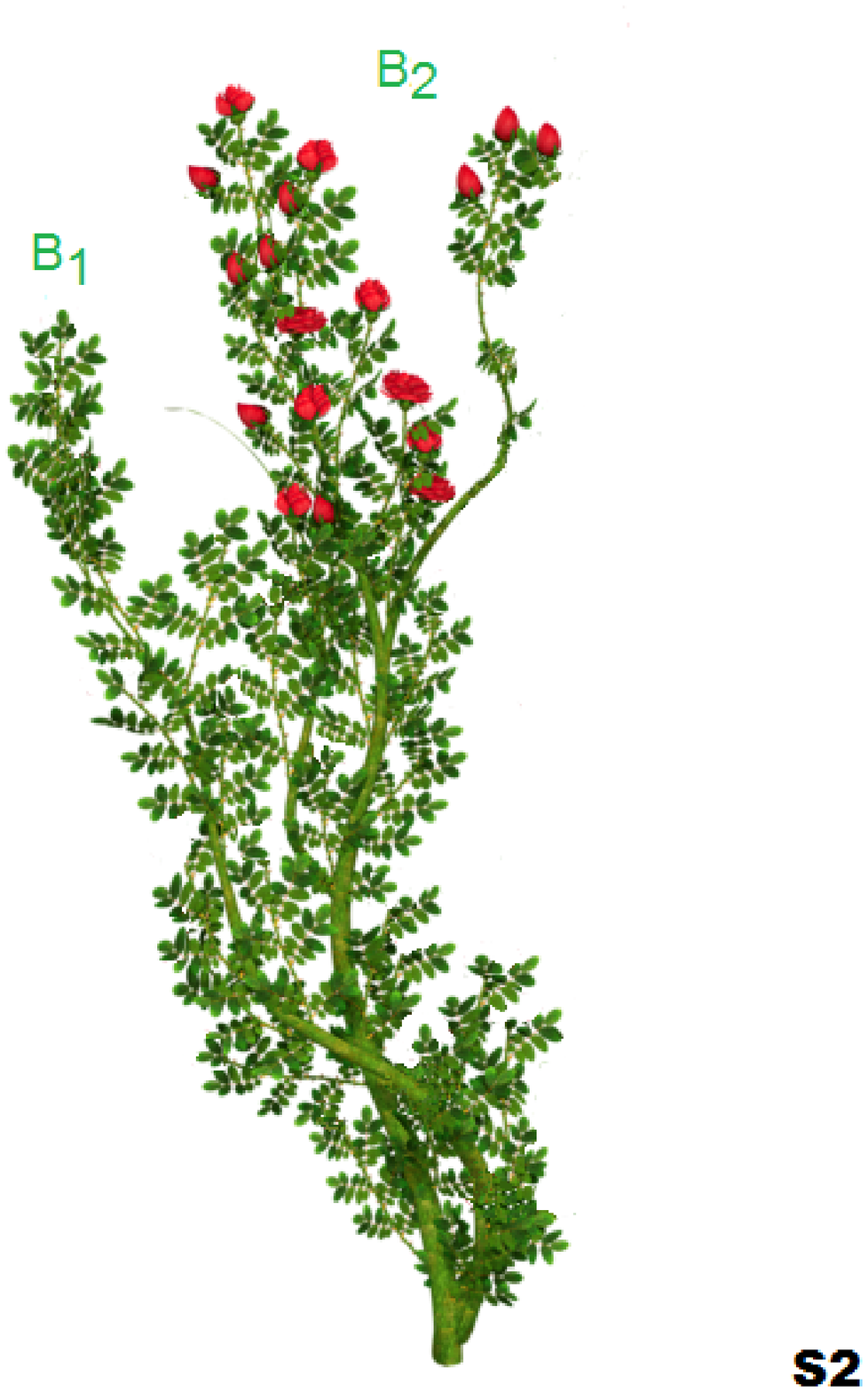} \includegraphics[width=4.9cm]{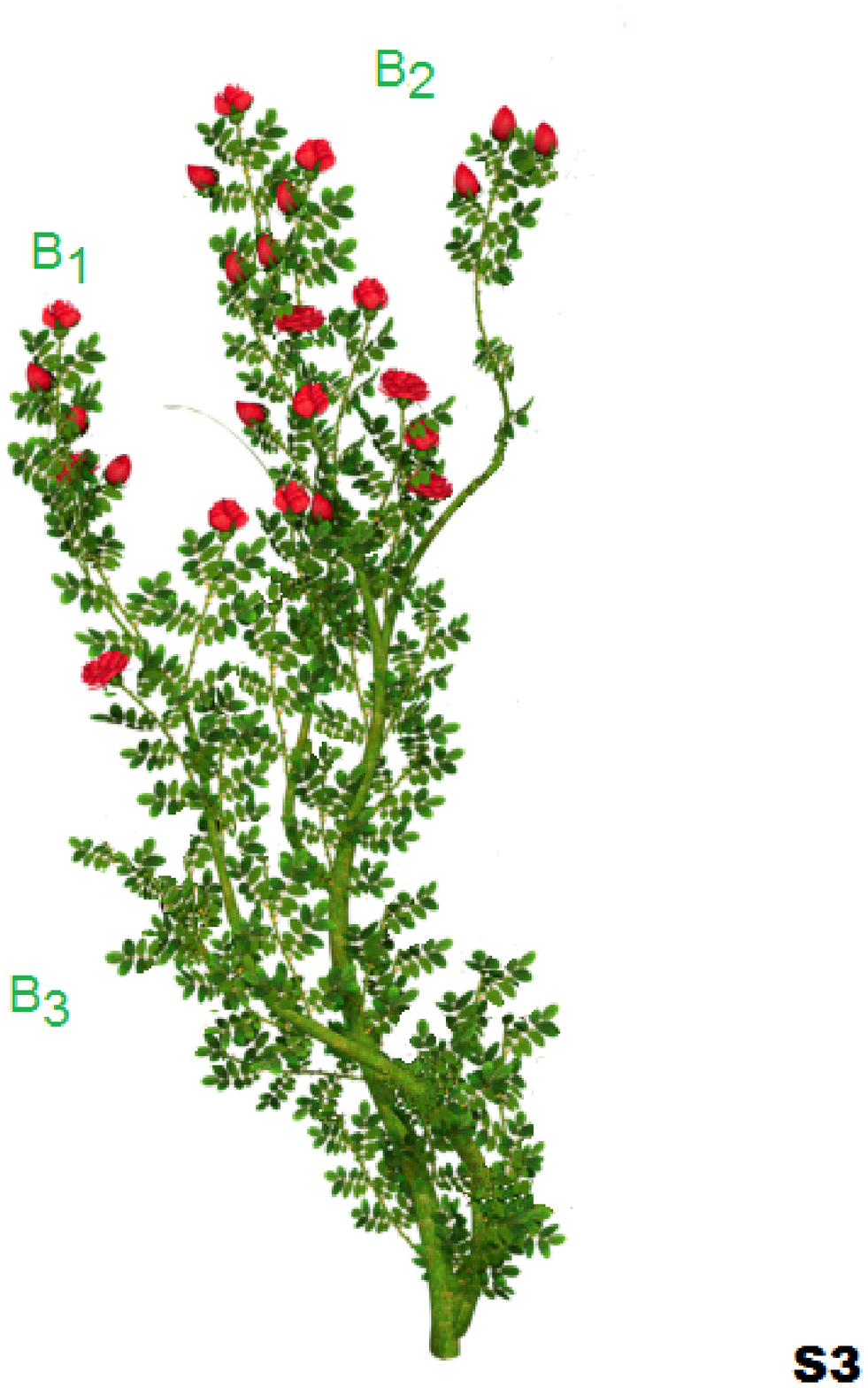}
\includegraphics[width=4.9cm]{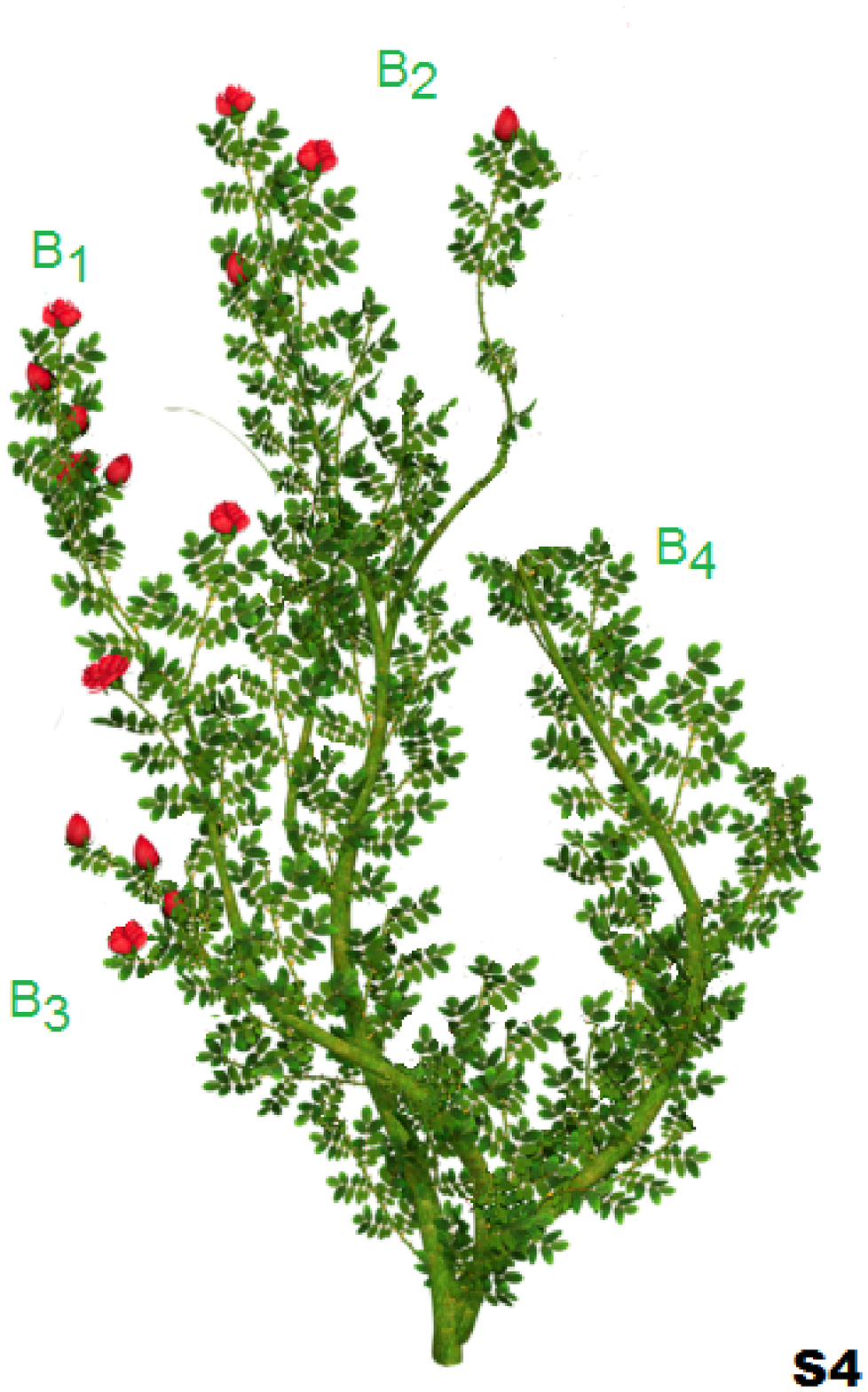}
\includegraphics[width=4.9cm]{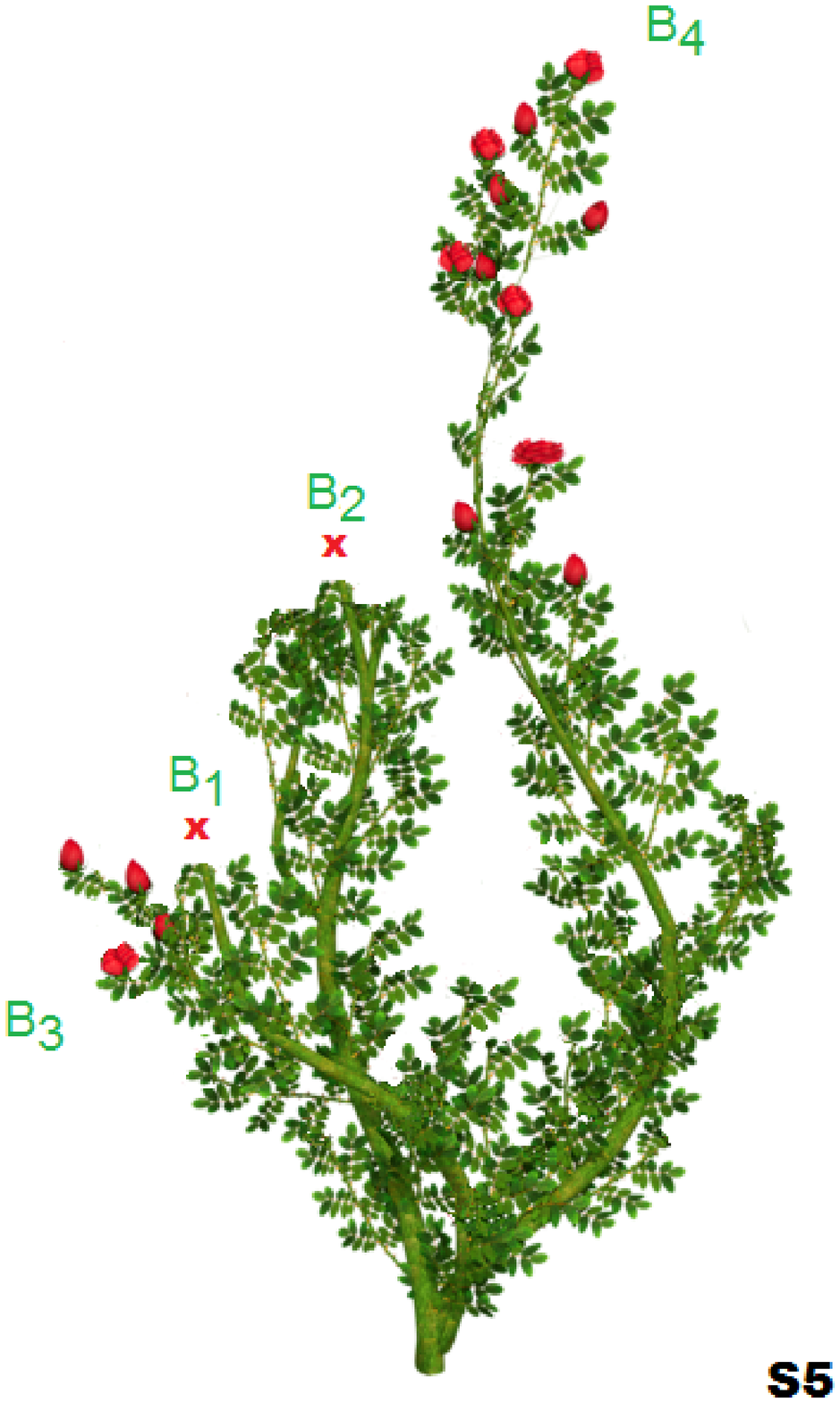}
\includegraphics[width=4.9cm]{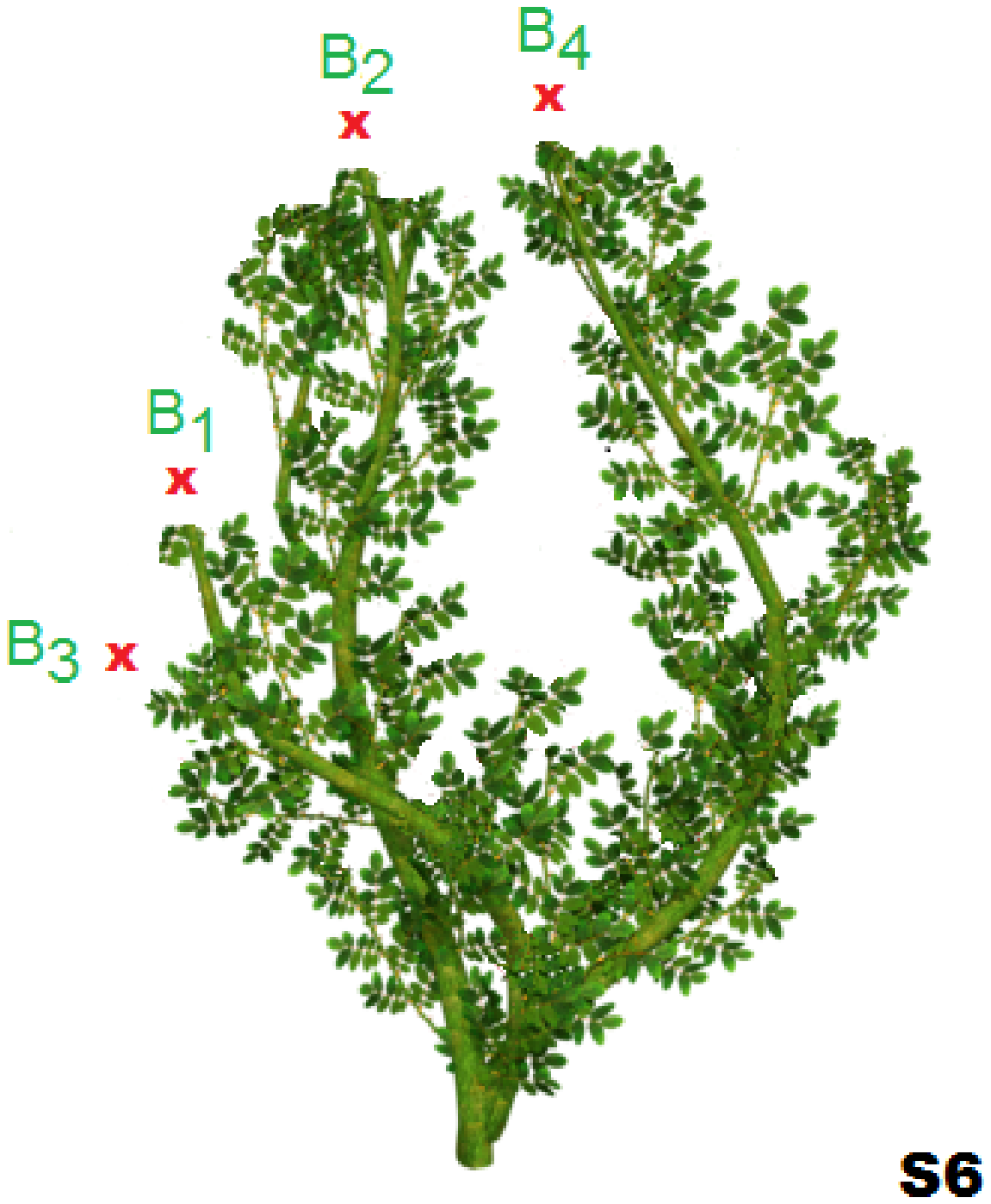}
\bigskip
\caption{\scriptsize{Example of six flowering stages (S1--S6) of a reblooming rosebush along the year (credit: free clipart) in which the evolution of four branches are represented (B$_1$, B$_2$, B$_3$, B$_4$).}}
\label{FigRosebush}
\end{figure}

On Figure \ref{FigRosebush} above, a simulation of the flowering stages of a reblooming rosebush along the year is given. Through this example, we aim to highlight the fact that merely counting flowers over time on a plant is not always relevant to study its flowering process. The six stages are described thereafter.
\begin{itemize}
\item S1 : before its first instant of flowering, the rosebush has developed two branches (B$_1$ and B$_2$).
\item S2 : branch B$_2$ gives an abundant flowering.
\item S3 : while B$_2$ is still blooming, branch B$_1$ produces a scattered flowering and a small branch B$_3$ is appearing at the root of B$_1$.
\item S4 : branch B$_3$ immediately comes into flower whereas B$_2$ is withering, and a sizable branch B$_4$ is growing from the base of the rosebush.
\item S5 : branches B$_1$ and B$_2$ are pruned after their 	whole decline while B$_4$ displays a sudden flowering.
\item S6 : at the end of the flowering process, the rosebush is pruned.
\end{itemize}

\bigskip

\begin{figure}[H]
\centering
\includegraphics[width=8cm]{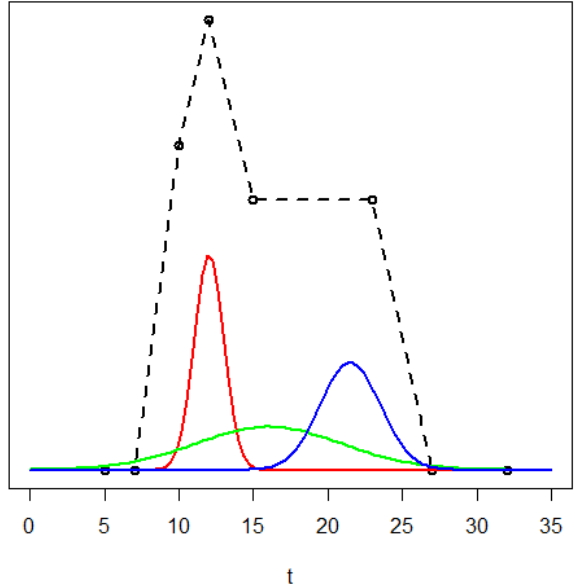}
\bigskip
\caption{\scriptsize{Shape of the simulated flowering curve of the example above (black dotted line), the waves detected by the GMM algorithm are coloured.}}
\label{FigRosebushSimuGMM}
\end{figure}

\noindent The gap between the flowering of B$_1$ and B$_2$ explains the spread of the first flowering event, common to both once-flowering and reblooming roses. Flowering of B$_3$ and B$_4$ corresponds to reblooming events, only observable in reblooming roses. On Figure \ref{FigRosebushSimuGMM}, we have represented the shape of this simulated flowering curve, to show that a naive evaluation of the number of flowers on the plant along the year clearly leads to conclude that this rosebush is once-flowering. The superimposition of the reblooming waves detected by GMM allows the opposite conclusion which is biologically more relevant.

\newpage

\nocite{*}

\bibliographystyle{acm}
\bibliography{GMMFlowering}

\vspace{10pt}

\end{document}